\documentclass[aps,graphicx,showpacs]{revtex4}  %% Journal Style
\def\bra{\langle} \def\ket{\rangle}
\usepackage{graphicx}

\begin{document}

\vbox{\hbox to \textwidth{\small \ Commun. Theor. Phys. 45 (5) (2006) 825-844 / Arxiv:
quant-ph/0512120 \hfill{\ } \hfill\hbox to 1cm{\hfill\thepage\hspace
{2mm}}}\vspace{1.6truemm} \hspace{2mm}}

\title{The General Quantum Interference Principle and the Duality Computer}
\author{Gui Lu Long}
\address{
Key Laboratory For Quantum Information and Measurements and
Department of Physics, Tsinghua University
Beijing 100084\\
Key Laboratory of Atomic and Molecular Nanosciences, Tsinghua
University, Beijing 100084, P. R. China}

\begin{abstract}
In this article, we propose a general principle of quantum interference for quantum
system, and based on this  we propose a new type of computing machine, the duality
computer, that may outperform in principle both classical computer and the quantum
computer.
 According to the general principle
of quantum interference, the very essence of quantum interference is the interference
of the sub-waves of the quantum system itself. A quantum system considered here can be
any quantum system: a single microscopic particle, a composite quantum system such as
an atom or a molecule, or a loose collection of a few quantum objects such as two
independent photons.
  In the duality
computer, the wave of the duality computer is split into several sub-waves and they
pass through different routes, where different computing gate operations are performed.
These sub-waves are then re-combined to interfere to give the computational results.
The quantum computer, however, has only used the particle nature of quantum object. In
a duality computer, it may be possible to find a marked item from an unsorted database
using only a single query, and all NP-complete problems may have polynomial algorithms.
Two proof-of-the-principle designs of the  duality computer are presented: the giant
molecule scheme and the nonlinear quantum optics scheme. We also proposed thought
experiment to check the related fundamental issues, the measurement efficiency of a
partial wave function.

\end{abstract}
\date{Received 15, Dec. 2005; Revised Feb. 6, 2006}
 \maketitle

%\tableofcontents
%\newpage
\section{Introduction}
\label{s1}

The particle and wave nature of microscopic object,  is one of the
most important aspect of the complementarity principle \cite{r1}.
A microscopic object possesses  both the attribute of a particle
and the attribute of a wave, and it displays one attribute at some
instances and the other attribute in some other instances. The
particle-wave duality of microscopic object is a very mysterious
feature of quantum mechanics. This is well stated by Feynman in
his physics lecture notes \cite{r2}
 \begin{quote}
"The quantum interference phenomenon is impossible, absolutely
impossible, to explain in any classical way, and it has in it the
heart of quantum mechanics",  "In reality, it contains the only
mystery".\end{quote}

Classically, waves behave quite differently from particles. A wave travels, and  splits
into many parts when it passes through slits in a wall. When these wave parts meet
again, they join together and interfere, constructively or destructively. If we make
some changes in one path, say, change its phase, then the interference pattern will
change. In contrast, a particle is something that has a definite position in
space-time. These seemingly contradicting properties are all exhibited in the very same
microscopic object. Taking the double-slits experiment as an example. Suppose an
electron starts from a source $S$, "passes through" a wall with two slits and then hits
on a screen. An interference pattern will appear on the screen after sufficient number
of particles have been fired. The quotation mark is a reflection of the fact that we do
not yet know whether a particle passes through only a single slit, or through both
slits {\bf simultaneously}. What we know is the fact that a particle has passed through
the wall and hit on the screen. Any attempt to  locate which slit a particle has passed
through will eliminate the interference pattern on the screen as shown in the which-way
experiment \cite{r3}. However without bothering to know which path it passes through,
the interference pattern shows up. By varying the phase differences between the two
paths, the interference pattern on the screen
 will change accordingly and in a determined
manner.

Interference of light through a double slits was first observed by Young \cite{r4}, and
it has played a vital role in the acceptance of the wave nature of light. With the
development of modern technology, interference of electrons  has been
obtained\cite{r5,r6,r7}, and the intensity of the electron can be made so weak that one
can observe only a single click on the screen at a given time \cite{r7}. A click on the
screen is a reflection of the particle nature of electron. Field-theoretic treatment of
photon interference was given by Walls in Ref.\cite{walls}.

 Dirac made a famous statement \cite{r8}: "
 {\bf Each
photon then interferes only with itself. Interference between two
different photons never occurs.}"  However, sometimes interference
 of two photons, for instance in
Ref.\cite{r9} in which two photons, each from the same single photon source but at
different instants, do interfere. It seems that Dirac was wrong. We will see that these
phenomenon can be naturally explained in the general interference principle of quantum
mechanics that we are going to present.  From this general principle of quantum
interference, interference can occur ont only in a single quantum system, but also
occur in composite quantum systems such as electrons in an atom, and even in a quantum
system with particles loosely distributed without any binding. This principle will
explain different interference phenomena in the quantum world in a unified way.

Physics and computation are closely related. A computing machine is any physical system
whose dynamical evolution takes it from one of a set of "input" states to one of a set
of "output" states.  It had been long considered that any computing process can be
regarded solely as a mathematical process regardless of the specific computer
desgin\cite{r10,r11}. This is  true for classical computers and it is explicitly stated
in the so-called "Church-Turing hypothesis"
\begin{quote}
Every `function which would naturally be regarded as computable'
can be computed by the universal Turing machine.\label{e1.1}
\end{quote}
Classical computing process is usually irreversible. Landauer
stressed the importance of irreversibility in classical computer
and  gave the minimum amount of energy a classical computer has to
dissipate in erasing or throwing away a bit of
information\cite{r12}. But Bennett showed that classical
computation can be made reversible\cite{r13}. Benioff has
constructed a reversible computer using explicitly quantum
dynamics\cite{r14}. Feynman pointed out that it was difficult for
classical computers to simulate quantum systems efficiently and
proposed to simulate quantum systems using quantum
computers\cite{r15}.

Deutsch replaced the subjective notions such as "would naturally
be regarded as computable"  in the Church-Turing hypothesis with
objective concepts, and reformulated it into the Church-Turing
principle\cite{r16}
\begin{quote}
Every finitely realizable physical system can be perfectly
simulated by a universal model computing machine operating by
finite means.
\end{quote}
Deutsch further pointed out that the classical universal Turing
machine does not satisfy the Church-Turing principle. A universal
quantum computer {$\cal Q$}, however, satisfies this Church-Turing
principle.

Thus it began to show that the  power of a computer depends closely on the realm of the
underlying physics principle. We may conjecture that within a given realm of physics,
all computing machines operating with the laws of that realm of physics are equivalent.
It is true that all classical computers are equivalent, any classical computer can be
perfectly simulated by a universal Turing machine with at most a polynomial slowdown.
It is also true that all quantum computers are equivalent \cite{r17}. However computers
in different realm of physics are not equivalent. For instance a computing machine in
the classical physics realm is not equivalent to a computing machine in the quantum
mechanics realm. A classical Turing machine can not simulate a quantum system
efficiently, whereas a quantum computer can. But among the quantum computers, they are
equivalent. The dominance of classical computations has led people to restrict their
thinking on the classical computation, and think that computation is solely a
mathematical process.

It should be stressed that the nature of a computer should be
judged not only by the realm of physics upon which it is
operating, but also on the extent to which it has exploited the
physics principles of that realm of physics.  An electric
classical computer working with transistors will be considered the
same as a computer working with mechanical means because their
underlying physics principles are the same.  We will show in this
article  that the quantum computer uses only the particle nature
of quantum system, but not the full power of quantum mechanics. It
is superior to classical computer because it has used the
superposition principle of quantum mechanics and this has enabled
the quantum parallelism in quantum computers.  To be exact, the
present quantum computer should have been called {\bf quantum
particle computer}, and the type of computer we are proposing here
should have been called quantum computer. For historical reasons,
we will follow the tradition to call these quantum particle
computer as quantum computer and call the computing machine we are
proposing as {\bf duality computer(DC)}. Incidentally, the basic
information carrier in the duality computer is called duality bit
or {\bf\it dubit} for short.

When talking about computational complexity, the NP-complete problem is one benchmark.
Problems such as the SATISFIABILITY problem \cite{r18} requires steps of computation
that scales exponentially with the size of the input. In classical computing theory, it
is not known if NP=P is true or not. However it is widely anticipated that NP-complete
problem is unlikely to have a polynomial solution. In quantum computer, no polynomial
algorithm has been found for any NP-complete problem so far. It is also anticipated
that $NP$-complete is not equal to $P$ in quantum computers either.

Duality Computer allows more powerful computing operations than
what are allowable for a quantum computer. The essential idea of
the DC is that the multi-dubits quantum system is treated wholly
as a single quantum system. Usually internal degrees such spins
are used as dubits. The duality nature of the whole system allows
us to divide the quantum wave of the system into multiple parts,
or paths, just like an electron passes through a double-slits and
divides its wave function into two parts. We can perform different
gate operations on different parts of the wave. This is a new type
of parallelism, the duality parallelism. Then the different
sub-waves are recombined to form the wave of the duality computer.
The result of the computation is then obtained by making a
measurement.
 The duality nature of quantum system may offer enormous computing power.
 The long-term
neglect of the  duality property in computation may be due mainly to
the lack of understanding of the quantum interference of quantum
systems with composite components.  The quantum duality, or
the particle-wave duality,  of a single microscopic object is well
accepted and understood, but the quantum duality of a composite
quantum system is less well understood. This lack of understanding
has hindered us from searching more powerful computing machines.

Associated with the quantum duality nature, we also examine other aspects in quantum
information processing, for instance the no-cloning theorem \cite{r19}.  With quantum
duality, though we cannot clone an unknown quantum state,  we can  divide the wave of a
quantum state into many sub-waves and each of the sub-wave carries exactly the same
unknown (internal) quantum state. Quantum state division operation is a different
operation from cloning, and thus it does not contradict the no-cloning theorem.

The article will be organized as follows. In section \ref{s2}, we will present the
general  quantum interference principle.  This includes a reformulation of the quantum
interference principle and the generalization to composite quantum systems. In section
\ref{s3} we will give a brief review of classical computer and quantum computer. In
section \ref{s4}, we will present a new operation available for duality computing:
quantum wave division and quantum wave combination. In section \ref{s5}, we present the
major construction of the duality computer. We point out that duality computer provides
with us duality parallelism that may enable duality computer to surpass quantum
computer. In section \ref{s6}, we will give, as proof-of-the-principle, two designs of
the duality computer: the giant molecule scheme and the nonlinear quantum optics
scheme. In section \ref{s7}, we give  duality algorithms for the unsorted database
search problem and NP-complete problems. We show that NP-complete = P may be true in
duality computer. In section \ref{s8}, a concluding remark is given.

\section{The general quantum interference principle}
\label{s2}

There are two types of interferences in nature: classical
interference and quantum interference. In classical physics, the
interference of sound waves and electromagnetic waves are such
examples. In these interferences, it is the effect of action that makes
the interference.  In quantum mechanics, the double-slits interference of
electrons is a typical example. It seems that in classical
physics, waves from different source can interfere as long as the
two waves have a definite phase difference. However, in quantum
mechanics, interference takes place when waves of the same quantum
object meet in space and time.  It is the interference of the existence of the
quantum system in quantum interference. Though these two kind of interferences are
related, they are different.  We restrict ourselves to quantum interference only
in this article.

Interference phenomenon has been an important stimulus  in the
development of quantum mechanics. Not only light, but also
particles such as electron and atom exhibit interference when
passing through a double-slit. The double-slit experiment is the
most popularly known experimental foundations of quantum
mechanics. Feynman paid great attention to the experiment, and
pointed out the essence of quantum mechanics lies in the
double-slits experiment.

Here we first give the general quantum interference principle. This principle is partly
a summary of already known results, and partly our generalization.
\begin{enumerate}

\item Quantum interference can happen for a quantum system. The
quantum system maybe one of the following categories: a single
particle such as an electron or fundamental particle, or a
composite system with several constituent particles such as an
atom or a molecule, or just a collection of particles with no
interaction, or binding at all, such as two independent photons.
We call the first two categories as a tight quantum system(TQS),
and the latter as loose quantum system(LQS).

\item Quantum systems, TQS or LQS, interfere only with themselves
through the interference of the waves describing the whole quantum
system. If the wave describing the whole quantum system is divided
into more than one path and then recombines, quantum interference
may occur.

\item Interference of the waves describing the whole quantum
system takes place when the waves coincide in the same region in
space and in the same period in time. In other word, interference
takes place at the same space-time location.   The total wave
describing the whole quantum system is the sum of all the
sub-waves.

\item Interference takes place when the waves are
indistinguishable in space and in time. If the system involves
identical particles, this has to be taken into account.

\item The wave of the quantum system is characterized by not only the center of mass
motion quantum number such as the wave number, but also the internal degree of freedom
of the quantum system such as polarization, spin, isospin and other quantum numbers of
the constituent particles etc. If the wave function for other degrees of freedom are
identical in all the sub-waves, then we can simply consider the center of mass
travelling wave functions.
\end{enumerate}

The general principle of quantum interference can be explained from Dirac's path
integral formalism. Suppose the wave function of a quantum system with $n$ constituent
particles, loose or bound, is $ \psi(x_1,x_2,\cdots,x_n;s_1,s_2,\cdots,s_n,t) $ where
$x_i$ and $s_i$ are the coordinate and the internal degree coordinate of the $i$-th
particle, then by generalizing the path integral to include also the internal degree of
freedom, we have
\begin{eqnarray}
\psi(x_1,\cdots,x_n;s_1,s_2,\cdots,s_n,t)= A\sum_{all\;paths}e^{{i\over \hbar}
S[x,s,x',s']}\psi(x_1',\cdots,x_n;s_1',s_2',\cdots,s_n',t'),
\end{eqnarray}
where $S[x,s,x',s']$ is the action along a path, and $x$, $x'$, $s$ and $s'$ are short
notation for all the $n$ variables respectively. It should be noted that a path here
refers to a path in which all the $n$ constituents move from their initial position to
their respective final positions.  The path integral can be performed in groups
separately so that the summation over all paths can be written as a summation over a
limited number of distinct paths. For instance, in a one-dimensional free particle
case, though there is only a single path from position A to another position B, the
electron can still assume numerous kinds of motion along that path: at each point in
the one-dimensional space, an electron can have different velocity. After performing
the summation over the different paths, we are left with a free particle wave function.
For a double-slits, such a treatment leaves us two distinct routes, each being a free
particle wave function, one from the upper path and one from the lower path.

Sometimes we also refer the wave as the probability amplitude
according to its interpretation in quantum mechanics. Let' study
these five criteria for quantum interference one by one. The first
criterion is a generalization of Dirac's statement into systems
with more than one microscopic particles. There are experimental
evidence. There is no doubt that a photon or an electron can only
interact with itself. The quantum interference of a composite
quantum system have also been experimentally demonstrated by the
Vienna group using giant molecules and biological large molecules
\cite{r31}. The interference of TQS is generally accepted, and one
usually has no difficulty in comprehend it.

The quantum interference of two-photon that are generated from type-II down conversion
simultaneously can occur \cite{r21}. The two photons produced in the type-II down
conversion was suggested by Shih as a "two-photon" particle, which should be treated at
the same footing as a composite particle though they are separated in space. Other two
photon interference phenomenon can also be explained in the "two-photon" picture
\cite{zhangsun}. A simplified illustration is shown in Fig. 7 of Ref.\cite{r21}. In the
middle is a nonlinear crystal. Due to type-II down conversion, two photons are
generated simultaneously. The frequency of these two photons are not necessarily equal.
Here
 One photon, is moving to the left, and after passing through a
 double-slits screen is detected
by detector D$_1$.  The other photon  is moving to the right and is detected by
detector D$_2$ after passing through a single slit in screen S$_2$. When observing the
coincidence measurements of the two detectors, interference pattern is observed in the
plot of the coincidence number versus the vertical position of detector D$_2$. This
interference phenomenon can be understood in the following way. The two photons can be
generated in several different possible positions in the nonlinear crystal. For the two
photons measured by the coincidence measurement, they can be generated in a position in
the upper part of the crystal, or a position in the lower part of the crystal. Thus two
waves, or sub-waves to be more exact, that describe the same two-photon system exist.
After generation they travel to the left and right respectively either through the
upper path, or through the lower path. At the detectors, the two sub-waves from these
two paths are indistinguishable at the detectors, that is the detectors can not
distinguish either the two photons detected are generated in a position at the upper
part of the crystal and travel leftwards and rightwards towards the them respectively
or from a position in the lower part of the crystal and then travel to them through the
lower path. Hence they combine and produce interference. It is really the interference
of the same quantum system, the system with two photons generated in the type-II down
conversion. There is no interference between the left photon and the right photon.
These two photons is a generalized composite particle system just like an atom or a
molecule. Different from a usual composite particle such as an atom, the constituent
particles such as the two photons are not combined together, and the distance between
them is changing. Whereas in an atom, the constituent particles are combined together
and are located in the same area in space.

It is also interesting to point out that the interference pattern
can also be observed if a delayed measurement is performed in the
right detector D$_2$. D$_2$ can be placed in a position that is
farther away from the crystal than that between D$_1$ and the
crystal. Hence the left photon is detected first, and the right
photon is detected later. There is no interference in the plot of
the counts in detector D$_1$ versus the vertical position.
However, if the coincidence of the two detectors is made, with
suitably chosen delay time according to the geometry of the
experimental setup, interference pattern is seen
clearly\cite{r21}.

The second criterion is a concretization of Dirac's statement for
a single photon in terms of wave functions and a generalization to
quantum system with multiple particles.   For a single quantum
particle such as a photon, the interference takes place only by
itself through its wave functions or amplitude probabilities. For
instance, when it passes through a double-slits, the wave function
of a photon is divided into two parts, and they then recombine at
the spot on the detection screen. The wave function of the photon
is a function of both the position, and  the photon polarization.
Suppose the two slits in the screen are identical in every
aspects, then the quantum wave function $|\psi\ket|\chi\ket$ will
be split into two parts with equal coefficients,
$${1\over 2}|\psi_u(t)\ket|\chi\ket={1\over 2}|\psi_u(t),\chi\ket,$$
in the upper path and
$${1\over 2}|\psi_d(t)\ket|\chi\ket={1\over 2}|\psi_d(t),\chi\ket$$
 in the lower path where the
coefficients have been renormalized to give the right intensity.
$|\chi\ket$ represents the internal wave function,  polarization
wave function, of the photon. The internal wave function has the
same form in the upper and lower paths because we have assumed
that the two paths do not affect the internal wave function. This
degree of freedom is usually ignored in double-slits experiment
with photons. In the neutron interference experiment however, this
degree of freedom has been taken into consideration
explicitly\cite{r22}.

  The difference
between the two kinds of complex systems, the TQS and LQS, is that
for the TQS such as a giant molecule, the whole system as a whole
moves like a single particle,  for instance when passing through a
double-slits, there are two possibilities, either from the upper
slit or from the lower slit, while for a LQS with two independent
photons, they do not have this combined behavior while in motion,
and there will be four possibilities when the two photons pass
through a double-slits, either both pass through the upper slit,
or the lower slit, or photon A passes through the upper and the
photon B passes through the lower slit, or the photon A passes
through the lower slit and the photon B passes through the upper
slit. As long as they satisfy the conditions for interference,
quantum interference will take place. For the TQS case, one can
still observe an interference pattern just like that for an
electron passing through a double-splits. However, for the LQS
such as two independent photons generated simultaneously,
interference can only be observed with two detectors and with
coincidence, and the interference pattern is the result of four
sub-waves.

It is interesting to generalize the double-slits into multiple-slits. When there are
$d$-slits on the wall, a quantum wave passing through the wall will be divided into $d$
sub-waves. We assume that these slits are identical, then at positions near the slits
just after the wall, each sub-wave is ${1\over d}|\psi\ket$ apart from a phase
difference in the coordinate wave function and $|\psi\ket$ includes both the spatial
wave function and the internal wave function. Here again a renormalization of the
coefficients has been taken so that if the $d$ sub-waves recombine with the same phase
at the detector, they will add up constructively to give $|\psi\ket$. For instance a
three-slits wall will divide a wave into 3 parts, each with $1/3|\psi\ket_i$ where the
subscript $i$ indicates the path number. However this information should not be
available at the detector, otherwise the interference pattern will disappear. If one of
the slits, say slit 1 is distinguishable from the other two slits, the sub-wave on path
1 becomes $\sqrt{1/3}|\psi\ket_1$ whereas the other two sub-waves becomes
$1/2\sqrt{2/3}|\psi\ket_2$ and $1/2\sqrt{2/3}|\psi\ket_3$ respectively. Then on the
screen, we should have a combined result of the incoherent summation of a single-slit
electron pattern having 1/3 of the total intensity together with an interference result
from slit 2 and slit 3 having 2/3 of the total intensity. This can be realized if we
let the spin polarization at slit 1 in spin up state and at slits 2 and 3 in spin down
state. If we have all the path information, then there will be no interference at all,
just like what happens in the which-way experiment.

Criterion 3 is the necessary condition for interference, just like
that for classical waves.  A wave train has a limited length,
which maybe different for different systems. For instance, for the
down-conversion produced photons\cite{r23}, the wave train is
about  16$\mu m$ long and has a time span of about 100$fs$.
Usually interference will be the strongest if all the waves
coincide closely in space and synchronize in time. If two
sub-waves do not overlap, then there will be no interference.

Criterion 4 further specifies the condition for interference. If the internal states
are different, they will not interfere even though they coincide in space and time. It
is also a special case of criterion 5 in which all other attributes of the sub-waves
are identical. The sub-waves become indistinguishable when all other degrees of freedom
become identical and the only difference between the different sub-waves at the
coalescing region are the amplitudes and the phases of the center of mass motion (or
called the spatial mode). If the other degrees of freedom are not identical between the
different waves, the overlapping between the different waves will decrease and hence
produces less or no interference. For instance, in a double-slits experiment, suppose
the sub-wave from the upper slit and the sub-wave from the lower path are made to have
different, orthogonal internal states, ${1/\sqrt{ 2}}|\psi_u,\chi_1\ket$ and ${1/\sqrt{
2}}|\psi_d,\chi_2\ket$. When they join together the total wave become $|\psi_t\ket={1/
\sqrt2}|\psi,\chi_1\ket+{1/ \sqrt2}|\psi,\chi_2\ket$. Then the probability to find a
particle is
\begin{eqnarray}
\bra\psi_t|\psi_t\ket&=&{1\over 2}\bra
\psi,\chi_1|\psi,\chi_1\ket+{1\over 2}\bra
\psi,\chi_2|\psi,\chi_2\ket+{1\over 2}\bra
\psi,\chi_1|\psi,\chi_2\ket+{1\over 2}\bra
\psi,\chi_2|\psi,\chi_1\ket\nonumber\\
&=&{1\over 2}\bra \psi,\chi_1|\psi,\chi_1\ket+{1\over 2}\bra
\psi,\chi_2|\psi,\chi_2\ket,\end{eqnarray}
 where $\bra
\chi_1|\chi_2\ket=0.$

A special case is the identical particle system. In the experiment in Ref.\cite{r9},
two photons emitted at different instants from a single quantum well are guided into
the two ports of a beam splitter. Then the two cases of one photon being reflected and
one photon being transmitted are indistinguishable. The two photons are identical
photons, though they are generated in different instants. If one could make the two
photons to have different frequency, then interference would have disappeared. The
interference of light from two independent, but synchronized lasers, can be understood
in this view. For instance, the light was made so weak that at any given instant with
high probability there was only a single photon in the experimental
set-up\cite{independent}. Because a photon produced from a laser is indistinguishable
from a photon from the other laser. These two waves can superpose at the detector to
interfere. It is somehow similar to the two-photon interference experiment where the
two photons can either be produced in the upper part or the lower part of the crystal.

Criterion 5  has been largely neglected before. In addition to the
space attribute, a quantum system also has other degrees of
freedom such as spin for electron, the polarization for photon,
the states of the electrons, nuclei of the atoms inside a
fullerene molecule. When a quantum system passes through two
identical slits, the wave function in fact has been split into
\begin{eqnarray}
{1\over 2}|\psi_u,\chi\ket,
\end{eqnarray}
for the upper slit, and
\begin{eqnarray}
{1\over 2}|\psi_d,\chi\ket,
\end{eqnarray}
for the lower slit. Here the internal states are identical. They
can become different if we apply different operations on the
internal degrees of freedom for the different parts. In fact, the
disappearance of interference in the which-way experiment is
because the internal state of the different paths are different
where the internal states in the different paths  are in fact
orthogonal.

Conversely, if we manage to make the spatial part of all the
sub-waves to have the same phase, when they combine, the spatial
part becomes a common factor and the summation is carried out only
in the internal part of the sub-waves. Then we will see
interference patterns arising from the internal degree of freedom.

In general, the total wave of a quantum system is the sum of all
the sub-waves from different paths, if the sub-waves from
different path are not orthogonal, then interference will occur,
whether partially or completely.

\section{Classical Computers and Quantum Computers}
\label{s3}

Though logic does not limit the set of computable functions of a
computing machine, physics does. In classical computer, all
possible computing functions are restricted by the set of
computable functions of the universal Turing machine. This
limitation is not due to mankind's wisdom in computer design, nor
by the ingenuity in constructing models for computation, but is
universal. This limitation can only be broken  by the use of a
different underlying physics, namely the kinds of operations that
are allowable in the computing process. Generally, low-level
physics is some approximation of a more advanced physics. For
instance, Newtonian mechanics can be regarded as the low speed
approximation of Einstein's relativistic mechanics, and classical
physics is the macroscopic limit of quantum mechanics. Computing
machines operating with an advanced physics principle will cover
the computing machines operating with a less advanced physics in
terms of computing power. In a universal quantum computer, quantum
parallelism exists, and it adds additional computing power over
classical computer. For instance, Shor's algorithm can factorize a
large integer in polynomial time\cite{r24}. Grover's algorithm can
find a marked item from an unsorted database with $O(\sqrt{N})$
steps\cite{r25,r26}, whereas $O(N)$ steps have to be searched on a
classical computer. The types of calculations available for a
quantum computer is also restricted by the physics principle, for
the unsorted database search problem Grover's algorithm is the
fastest that a quantum computer can do\cite{r27}.

Typically, a quantum computer consists of two components, a finite
processor and an infinite memory tape\cite{r16}. The computation
is performed in steps of fixed time duration $T$, and during each
step  the processor only interacts with  a finite part of the
memory, and the rest of the memory remains static. The state of
$\cal Q$ is a unit vector in the space $\cal H$ spanned by the
simultaneous eigenvectors
\begin{eqnarray}
|x;\hat{n};\hat{m}\ket\equiv|x;n_0,n_1,\ldots,n_{M-1};\ldots,m_{-1},m_o,m_1,\ldots\ket,
\label{e1.4}
\end{eqnarray}
where $x$ is the "address" number of the currently scanned tape location. $\hat{n}$ is
the state of the processor and $\hat{m}$ is the tape eigenvectors. This is the
computational basis states.  The dynamics of $\cal Q$ is represented by a constant
unitary operator $U$ on $\cal H$. $U$ specifies the evolution of any state
$|\psi(t)\ket\in \cal H$. The effect of $U$ on $\cal H$ can be represented by
\begin{eqnarray}
\bra x';\hat{n}';\hat{m}'|U|x;\hat{n};\hat{m}\ket
=\left[\delta^{x+1}_{x'}U^{+}(\hat{n}',m'_x|\hat{n},m_x)+
\delta^{x-1}_{x'}U^-(\hat{n}',m'_x|\hat{n},m_x)\right] \Pi_{y\ne
x}\delta^{m_y}_{m_y}, \label{e1.5}
\end{eqnarray}
which indicates that the scanning head can move one step to the
left or to the right each time. $\hat{n}'$ and $\hat{n}$ are the
labellings of the processor's internal state, and $m_x'$ and $m_x$
are the labellings of the tape that have been involved in the
interaction with the processor after and before the tape head
movement.

Two models of quantum computer exist. The universal quantum Turing
machine resembles straightforwardly the classical universal Turing
machine\cite{r16}. Similar to the von Neunman's logic circuit
network model of classical computer, there is the quantum circuit
network model\cite{r17}. Yao has showed that the two models of
quantum computer are equivalent\cite{r17}:
\begin{quote}
Let $\cal M$ be a quantum Turing machine and $n$, $t$ be positive
integers. There exists a quantum Boolean circuit $K$ of size
$poly(n,t)$ that $(n,t)$-simulates $\cal M$.
\end{quote}
A quantum Boolean circuit $K$ with $n$ input variables is said to $(n,t)$-simulate a
quantum Turing machine $\cal M$, if the family of probability distributions
$P_{\tilde{x}}$, $\tilde{x}\in \{0,1\}^n$ generated by  $K$ is identical to the
distribution of the configuration of $\cal M$ after $t$ steps with $\tilde{x}$ as
input. Yao also proved that a universal quantum Turing machine can simulate any given
quantum Turing machine with only a polynomial slow-down, and hence all quantum Turing
machines are computationally equivalent.

Quantum Turing machine model is attractive to computer scientists.
To physicists, the quantum computational network model is more
attractive. In quantum computational network, there are quantum
gates, sources, sinks and unit wires. A quantum computing process
can be viewed as a certain number of qubits prepared in some
initial states, passing through a series of quantum gates and
finally be measured. In a quantum computational network, the
evolution is from left to right. All quantum gate operations are
unitary. Unitarity is a requirement in quantum mechanics that all
observables are hermitian, especially the hamiltonian of the
system. A quantum gate operation can be seen as the controlled
evolution of the quantum system consisting of all the qubits.
Suppose that the initial state of a quantum computer is
$|\psi_i\ket$. After the action of a sequence of quantum gate
operations, the state of the quantum computer becomes
$|\psi_f\ket$, and the process can be written as
\begin{eqnarray}
|\psi_f\ket=G_nG_{n-1}\ldots G_1|\psi_i\ket, \label{e1.6}
\end{eqnarray}
where each $G_i$ is a gate operation. A gate operation can be an
arbitrary unitary operation on the Hilbert space $\cal H$  span by
the $2^n$-basis states of the $n$ qubits system. After the
completion of the gate operation a projective measurement on the
quantum computer is made to read out the result. Certain outcome
will appear probabilistically.

The most distinct feature of quantum computation is the quantum
parallelism. When acting on the superposition of all the
computational basis states with a unitary operation effecting a
function $f$, $U_f$,
\begin{eqnarray}
U_f{1\over \sqrt{N}}\sum_{i=0}^{N-1}|i\ket|0\ket={1\over
\sqrt{N}}\sum_{i=0}^{N-1}|i\ket |f(i)\ket, \label{e1.7}
\end{eqnarray}
the quantum computer calculates all the function values of the
function $f$ and puts them in the second register. The
right-hand-side of eq. (\ref{e1.7}) contains the result of an
arbitrarily large number $N$ computations of the function $f$,
which would require a classical computer $N$ number of
computations to complete.

Unitarity of operations in quantum computer has set very stringent
restrictions on the computing power of the quantum computer. For
instance, using the unitarity property, it has been shown that
Grover's algorithm is optimal for quantum computers\cite{r27} in
finding a marked state from an unsorted database.  Though great
computational powers have been delivered  by the quantum computer,
there are still limitations. There have been no polynomial
solutions to any NP-complete problems so far. In some problems,
the quantum computer offers no speedup at all. It seems that the
quantum computer maybe efficient in simulating local quantum
systems. But it is very restricted in solving general mathematical
hard problems. For example, the Grover algorithm only achieves a
square-root speedup, and it still scales exponentially with the
input size.

Can we do better than quantum computer in computations?  The power
of a computing machine is ultimately determined by the underlying
physics principle.  If we restrict ourselves to only classical
physics, then we can only have the computing power of a classical
Turing machine. Though there have been many efforts to find
polynomially quantum algorithms for NP-complete problems, it seems
difficult. We can do better if we go out of the underlying physics
realm of quantum computer.

\section{quantum wave divider and quantum wave combiner in duality computer} \label{s4}

We use quantum optics to illustrate some basic element in a quantum duality
information processer.
We use the two polarization states of a photon as the two quantum
levels of a dubit. A dubit possesses the duality nature of quantum
mechanics and thus can allow more operations than a qubit where it
is a special case of a dubit when a dubit takes on only a single
path. We denote the horizontal polarization state as $|0\ket$ and
the vertical polarization state as $|1\ket$. The photon
polarization can be rotated through a polarization modulator, for
example in a Pockel cell.  We denote the effect of such a polarization
modulator as $U_\theta$, mathematically, it can be expressed as
\begin{eqnarray}
U_\theta|0\ket&&=\cos\theta|0\ket+\sin\theta |1\ket,\nonumber\\
U_\theta|1\ket&&=-\sin\theta|0\ket+\cos\theta|1\ket.
\end{eqnarray}

A phase modulator can add a given phase to a photon state. It
is usually a variable wave-plate. Its effect can be written
as
\begin{eqnarray}
P_\lambda|\psi\ket=e^{i\lambda}|\psi\ket.
\end{eqnarray}

Here the effect is written out as if the dubit has only a single
path. When a dubit has two paths, the effects of such devices are
only on the specific sub-wave along a path.

\begin{figure}[h]
\begin{center}
\caption{A Mach-Zehnder Interferometer (a) without phase
modulator, and with a phase modulator (b).}
\includegraphics[width=10cm]{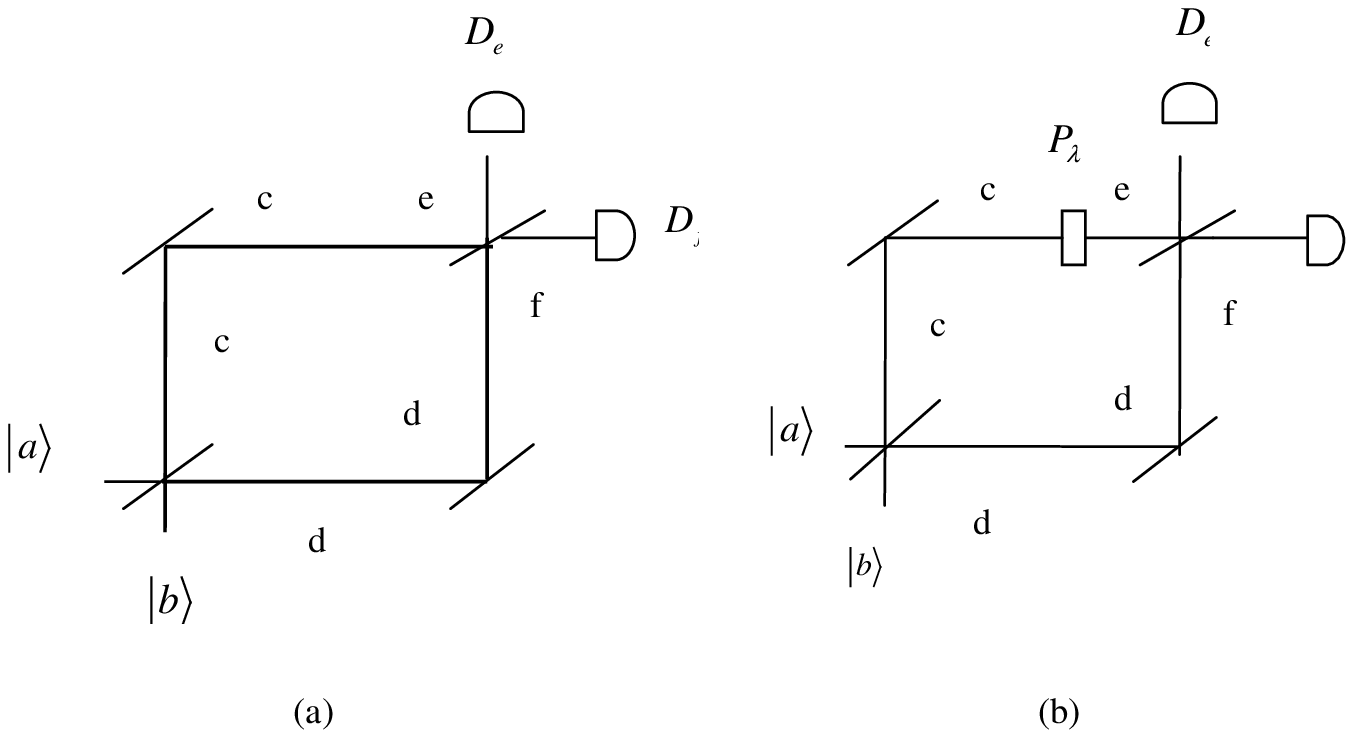}
 \label{f1}
\end{center}
\end{figure}

A beamsplitter is a common optics component. There are two kinds of beamsplitters: the
ordinary beamsplitter or beamsplitter for short, and the polarization beamsplitter. A
beamsplitter is insensitive to the photon polarization states, whereas the polarization
beamsplitter is sensitive to the polarization of the photon.  A beamsplitter with two
input modes $a$ and $b$ and two output modes $c$ and $d$ is shown in Fig.\ref{f1}a.
When a photon enters a beamsplitter from one port, depending on which side the photon
hits the beamsplitter, it transforms the photon state into
\begin{eqnarray}
|a\ket&&\longrightarrow{i}\cos\theta|c\ket+\sin\theta|d\ket,\nonumber\\
|b\ket&&\longrightarrow\sin\theta|c\ket+i\cos\theta|d\ket.
\end{eqnarray}
For a 50-50 beamsplitter, $\theta=\pi/4$. Then its transformation
is
\begin{eqnarray}
|a\ket&&\longrightarrow{i\over \sqrt{2}}|c\ket+{1\over
\sqrt{2}}|d\ket,\nonumber\\
|b\ket&&\longrightarrow{1\over \sqrt{2}}|c\ket+{1\over
\sqrt2}|d\ket.
\end{eqnarray}
It means that a 50-50 beamsplitter reflects  the
photon and transmits the photon with half probability respectively. The reflected
part has additionally a half $\pi$ phase factor: $e^{i\pi \over
2}=i$. Suppose a photon enters a beamsplitter from input mode $a$,
after the beamsplitter, the state of the photon becomes
\begin{eqnarray}
{i\over \sqrt{2}}|c\ket+{1\over \sqrt{2}}|d\ket.\nonumber
\end{eqnarray}
This resembles the case of a double-slits in which the path
information is known: we know the information of which path the
photon will take if we place a detector far away from the
beamsplitter on each side of the beamsplitter. In this case, the
two sub-waves of the output ports do not recombine so no
interference occurs between the two output mode of the
beamsplitter. To resemble a double-slits interference, we have to
eliminate the path information, and we need another beamsplitter
arranged in a Mach-Zehnder interferometer as shown in Fig
\ref{f1}b, then at the detector $D_f$ interference of the
sub-waves takes place.  The state of the photon goes the following
changes leading to the detector
\begin{eqnarray}
|a\ket\longrightarrow BS_1
\begin{array}{ccccc}
\nearrow & i\sqrt{1\over 2}|c\ket\rightarrow & M_1 \rightarrow &
i\sqrt{1\over 2}|c\ket
& \searrow\\
\searrow &\sqrt{1\over 2} |d\ket\rightarrow  & M_2\rightarrow
&\sqrt{1\over 2} |d\ket &\nearrow\end{array}
 BS_2\rightarrow \left(i\sqrt{1\over 2}\sqrt{1\over 2}+\sqrt{1\over
 2}i\sqrt{1\over 2}\right)|f\ket=i|f\ket.
 \end{eqnarray}
Here the two sub-waves interfere constructively at the detector
$D_f$. It is clear, the sub-waves from the two paths each
contributes $1/2$ to the total wave at the detector. To simulate
the double-slits experiment, we can vary the phase factor in one
of the arms as shown in Fig.\ref{f1}b, then the wave function
experiences the following change
\begin{eqnarray}
|a\ket\longrightarrow BS_1
\begin{array}{cccccc}
\nearrow & i\sqrt{1\over 2}|c\ket\rightarrow & M_1 \rightarrow
P_\lambda \rightarrow& ie^{i\lambda} \sqrt{1\over 2}|c\ket
& \searrow\\
\searrow &\sqrt{1\over 2} |d\ket\rightarrow  & M_2\longrightarrow
&\sqrt{1\over 2} |d\ket &\nearrow\end{array}
 BS_2\rightarrow i\left(1+e^{i\lambda} \over 2\right)|f\ket+
 \left({1-e^{i\lambda} \over 2}\right)|e\ket.
 \end{eqnarray}
As $\lambda$ changes from 0 to $\pi$ and $2\pi$, we have at the
detector a maximum, then a minimum and then another maximum, which
resembles perfectly a double-slits experiment. In detector $D_e$,
we see a complementary result. In
the following we will use interchangeably the double-slits and the
Mach-Zehnder interferometer to describe the physical realization
schemes of the duality computer.

 {\bf A quantum wave divider} is a device that divide a quantum wave
 into several sub-waves, and each sub-wave possesses
 the same internal quantum state as that of the input wave function.
 Let a particle passing through a
 double-slits, there are two
 paths: one from the upper slit and the other
 from the lower path. Sub-waves from these two paths will interfere at the
 screen where they become indistinguishable. In addition to the path information,
 a particle has also
 internal degrees of freedom, for instance a photon has polarization
 and an electron has spin. During the
 process of the particle passing through the double-slits, the
 internal degrees of freedom remains the same, say at  $(a|0\ket+b|1\ket)$, the only thing that
 is different is the spatial information. Hence the upper
 wave maybe written as
 \begin{eqnarray}
 |\phi_u\ket={1\over 2}(a|0\ket+b|1\ket)|\psi_u\ket,
 \end{eqnarray}
 whereas the low-path wave function is
\begin{eqnarray}
  |\phi_d\ket={1\over 2}(a|0\ket+b|1\ket)|\psi_d\ket,
\end{eqnarray}
where $|\psi_u\ket$ and $|\psi_d\ket$ are the spatial
wave function for the upper and lower path respectively.
It is interesting to note that here we have two sub-waves that has
exactly the same internal states. Their only difference is in the
spatial mode. It is not a clone of the state of one particle onto
another particle, {\bf it is a division of state of the same
particle}. When there are multi-slits in the wall, the quantum
wave of the particle will be divided into multiple sub-waves.
Again each sub-wave has the same internal states.

Now we generalize the above discussion to a complicated quantum system such as a
giant
molecule. Suppose a giant molecule has $n$ spin-1/2
nuclear spins. Each nuclear spin is used as dubit. Let the states
of these $n$ nuclear spins be denoted $|\chi_n\ket$ .
When the giant molecule passes through a
double-slit, the wave is divided into two sub-waves,
$|\psi_u\ket|\chi_n\ket$ and $|\psi_d\ket|\chi_n\ket$. The
internal state of these sub-waves are identical. This is a special
design of a quantum wave divider (QWD). We can generally assume that for
a general quantum system, there exists an QWD operation that divides the total
quantum wave into several sub-waves in different spatial modes but
with identical internal wave
function.
This is possible
when we use the quantum duality property. This important
ingredient is missing in a quantum computer.

\begin{figure}
\begin{center}
\caption{A quantum wave divider and a quantum wave combiner}
\includegraphics[width=10cm]{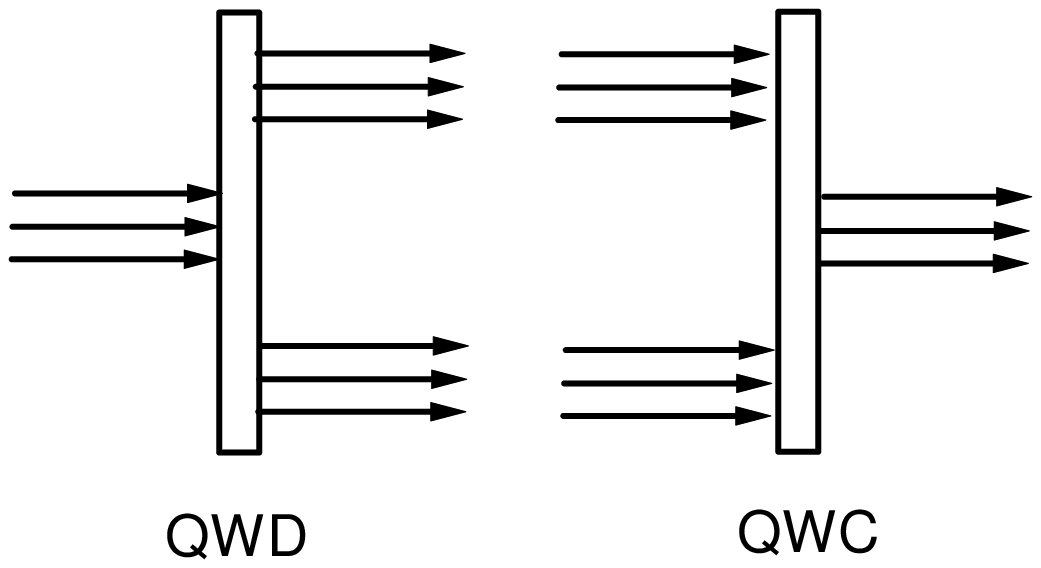}
 \label{f2}
\end{center}
\end{figure}

{\bf A quantum wave combiner(QWC)} is a device that does the reverse effect of a
quantum wave divider: it combines the sub-waves of a quantum system into a single wave.
For instance two sub-waves from a double-slits combine at the screen and interference
occurs. In general we assume such a device can be built for a given quantum system. QWD
and QWC are crucial for the duality computer. When a quantum system is not measured, it
behaves like a wave, can be divided or combined, and when it is measured it behaves
like a particle. In this article we represent the QWD and QWC using symbols as shown in
Fig.\ref{f2} where a quantum system with dubits are drawn.

Before going to a general DC, we demonstrate the use of a single
dubit to perform  duality computing. We write out here only the internal state of
the quantum system which is the state of the dubit. After the QWD, the state of a
dubit $a|0\ket+b|1\ket$ is divided into two sub-waves passing in
two different paths. On each path, we perform some gate operation,
which is composed of a series of unitary gate operations, on the
dubit. For instance we can perform only a unitary operation $U$ on
the upper path, and the upper sub-wave becomes
\begin{eqnarray}
 |\psi'_u\ket={1\over 2}U(a|0\ket+b|1\ket)={1\over
 2}(a'|0\ket_u+b'|1\ket),
 \end{eqnarray}
 where we have omitted the spatial part of the sub-wave.
At the screen the two sub-waves combine to form the  wave
\begin{eqnarray}
|\psi\ket_{f}={a+a' \over 2}|0\ket+{b+b' \over 2}|1\ket.
\end{eqnarray}
Here again the spatial wave function is not written out explicitly. It is always possible
to arrange the two paths to have the same lengths so that the phase difference between
the two paths are multiples of $2\pi$, namely at the QWC the spatial wave functions from the
two paths are identical and is a common factor.
 After measurement, the result is then read out.

Wootters and Zurek have shown that quantum state can not be
cloned\cite{r19}. In a cloning process, it is to clone the quantum
state of one particle onto another different particle. In this
splitting of quantum state in a double-slits, it is self division
of the quantum state on  different spatial degrees freedom,
however the whole wave, both the internal wave and the spatial
wave  together represent the state of the particle.

\section{The  Duality Computer and duality parallelism}
\label{s5}

A  duality computing machine uses the particle-wave duality
property of quantum mechanics for computation. In a duality
computer, the wave functions of the multi-dubit  duality computer
can split into two paths. The two paths have the same spatial length,
that is, when a duality computer passes through these two paths and recombine its
wave function at the QWC, the spatial mode of the two sub-waves
are in phase.  If there is only one path, then the duality
computer reduces to quantum computer which is presently under extensive
study worldwide.

Two components are crucial in  duality computer: the QWD and the
QWC, as have been introduced in previous section. For a single
particle quantum system, a QWD is just a beamsplitter or a
double-slits. When a photon passes it, there are 50\% probability
to be reflected and 50\% probability to be transmitted through the
beamsplitter. But for a multi-photon quantum system, a QWD is much
more complicated, and I shall give a specific implementation of a
QWD in later section. After the QWD, the wave splits into two
sub-waves in different paths. We can perform gate operations on
each sub-wave separately. Then the two sub-waves are combined in a QWC
so that interference may take place. The result of the calculation
can be read out by a final measurement. For a single quantum
particle, a QWC is just the screen in a double-slits experiment.
However for multi-photon system, the construction of QWC is
complicated and examples will be given in later section.

There are several issues I should like to emphasize.

 First, on a single path,
the gate operations is assumed unitary. Unitary operation is performed on a path in a
duality computer just as if we were operating on the whole quantum system. Dubit is on
constant motion, because particle wave duality is reflected when the quantum system is
moving. For a quantum computer, it is usually located at some area in space so that
when one makes a measurement he/she always find the quantum computer there. Though
there are some proposed physical realizations using flying qubits such as photons, it
is essentially a quantum particle computer as it is equivalent to the static quantum
computer realization. Therefore
 some modification on the way to
perform gate operation should be made. We will discuss this in the giant molecule
scheme in the next section. Though the operation in each path is unitary, the operation
on the whole duality computer is not. This is not surprising for we are considering
only a part of the space. For instance in the Mach-Zehnder interferometer setup, if we
just look at one outport we sometimes do not observe a photon though a photon is
injected into the interferometer. However if we look at both detectors we always
observe one when one photon is injected. This is fundamentally different from a quantum
computer where every gate operation must be unitary, and the whole quantum system takes
only a single path. Thus it is apparent that when there is only a single path in a
duality computer, it reduces to a quantum computer. The relationship between duality
computer, quantum computer and classical computer can be seen as follows: when a
duality computer is allowed to compute with only a single path, duality computer
reduces to a quantum computer. There is no duality in quantum computer, but
superposition and entanglement are still available in quantum computer. If we allow a
quantum computer to calculate using only the computational basis states, a quantum
computer reduces to a classical computer where neither superposition nor entanglement
are present.

Secondly, besides the QWD and QWC, other gate operations are
needed to implement the unitary operations in each individual
path. It has been proven that a set of basic gate operations is
sufficient to construct any unitary operation\cite{r28}. It is
known that single bit rotation gate and two qubit control
not(CNOT) gate form a set of universal gate\cite{r30}. This
universal set of gate can also be employed in duality computer.
Of course other sets of universal quantum gate operations could also be adopted.
As an example a universal set of  duality computing gate operations can be:
the QWD, QWC, 2-dubit CNOT and single bit rotations.

Thus the computing process in a quantum duality computer can be
illustrated as follows
\begin{eqnarray}
|\psi,\phi\ket\longrightarrow {\rm QWD}
\begin{array}{lllll}
 \nearrow & p_1|\psi,\phi_1\ket & \longrightarrow & p_1U_1|\psi,\phi_1\ket & \searrow     \\
 \searrow & p_2|\psi,\phi_2\ket & \longrightarrow & p_2U_2|\psi,\phi_2\ket  & \nearrow
 \end{array}{\rm {QWS}}\rightarrow
 (p_1U_1+p_2U_2)|\psi,\phi\ket,\label{double}
 \end{eqnarray}
 where $p_1\ge 0$, $p_2\ge 0$ and $p_1+p_2=1$, $\psi$, the first
 part in the state ket is the internal wave function which is the
 state of the duality computer, $\phi$ is the spatial wave
 function.
 The gate operation $U_1$ and $U_2$ are both unitary, and they act on
 the internal wave function. They
 themselves are composed of basic one bit and two bit gate
 operations. However, when they recombine at the QWS, which
 leads to
 \begin{eqnarray}
 p_1U_1|\psi,\phi_1\ket+p_2U_2|\psi,\phi_2\ket\rightarrow
({p_1U_1+p_2U_2})|\psi,\phi\ket,\label{duapar}
\end{eqnarray}
where the subscript in $|\phi_i\ket$ has been dropped because the spatial part of the
two sub-waves become identical at the QWC. The total operation is not unitary. For
instance for a symmetric QWD, $p_1=p_2=1/2$,
\begin{eqnarray}
\left(U_1+U_2 \over 2\right)^{\dagger}\left(U_1+U_2 \over 2\right)
&=&{1\over 4}\left(U_1U_1^{\dagger}+U_2
U_2^{\dagger}+U_1U_2^{\dagger}+U_2U_1^{\dagger}\right)\nonumber\\
&=&{1\over 4}\left(2I+U_1U_2^{\dagger}+U_2U_1^{\dagger}\right)\ne
I.
\end{eqnarray}

Eqs. (\ref{double}) or (\ref{duapar}) show that the duality
computer has a more powerful parallelism, the duality parallelism.
One can perform different  operations to the sub-waves in
different paths. While in quantum computer, this parallelism is
absent. Thus in duality computer, both the products of unitary
operations and the linear superpositions of unitary operations are
permissible. This contrasts to the quantum computer where only the
product of unitary operations are allowed. Since every $2^n\times
2^n$ matrix can be written as linear combination of unitary
matrices, therefore a  duality computer can implement any type of
operations in the Hilbert space.

This is the simplistic  duality computer model. Instead of dividing the quantum wave
into two sub-waves, we can divide it into multiple sub-waves. Furthermore, the division
can also be done for the sub-wave in each path to produce sub-sub-waves. In theory,
this division can be performed at an arbitrary level. This further division may provide
convenience and additional benefit in solving a specific problem.  It will be an
interesting investigation to study the computational complexities of the duality
computer. We can use the number of paths in a QWD and the levels of use of QWD to
describe the structure of a duality computer. For instance we call the DC described in
Eq.(\ref{double}) as a 1-level 2-paths  duality computer. If the divider has 3 outputs
where another QWD is  used at path 2 in Eq.(\ref{double}), then the duality computer
will be a called a 2 level 3-paths
 duality computer.

\section{Implementations of DC: the giant molecule scheme and the
nonlinear quantum optics scheme} \label{s6}

I give two physical realizations of the  duality computer. One is
the giant molecule scheme, and the other is the nonlinear quantum
optics realization.

\subsection{The Giant Molecule  Duality Computer} 
\begin{figure}[h]
\begin{center}
\caption{A Duality Computer with Nuclear Spins from a Giant
Molecule. }
\includegraphics[width=10cm]{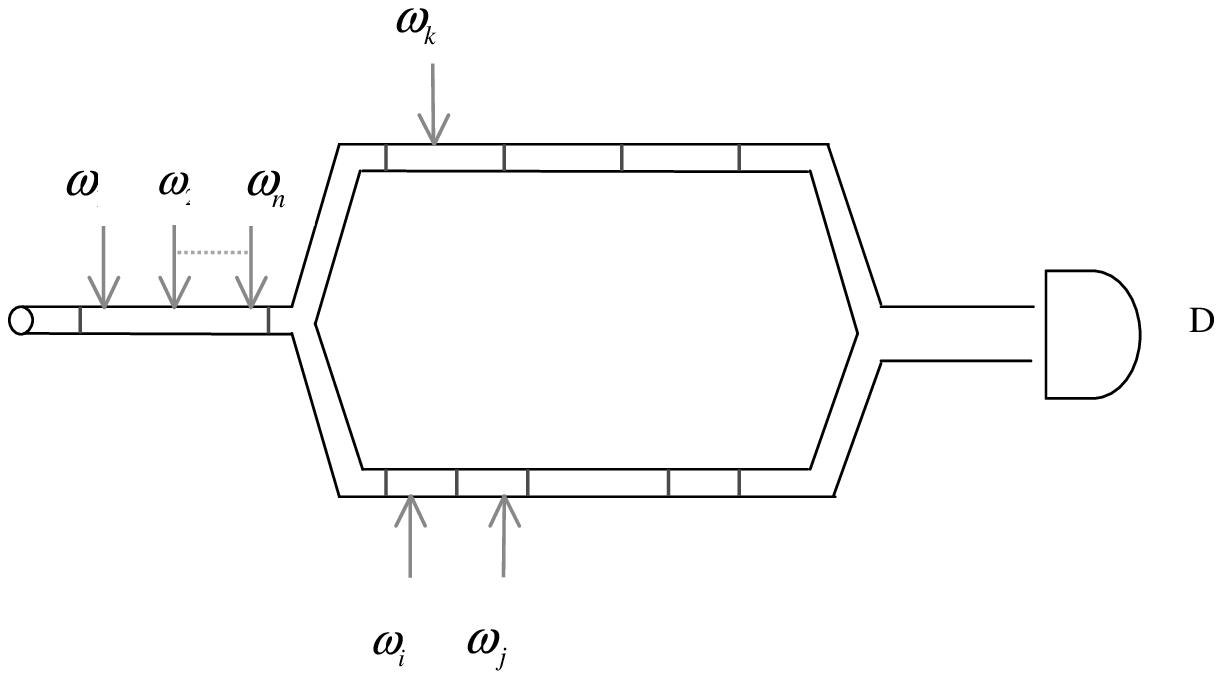}
 \label{f3}
\end{center}
\end{figure}

A
A spin-1/2 nuclear spin has two quantum state.   $n$
 nuclear
spins might act as an $n$-dubit duality computer. However if we
let these $n$ dubits to pass through a double-slits device, there
are many configurations. Say we have two dubits, then there will
be 4 possibilities for the two particles to be measured at a given
place on the screen: 1) both pass through upper slits; 2) both
pass the lower slits; 3)particle A passes through upper and
particle B passes through lower slit; 4) particle A passes through
lower slit and particle B passes through the upper slit. What we
need for the duality computation are the two paths that both
particles go through the upper slit or the lower slit. Hence it is
not workable if we simply use $n$ number of dubits. We need
something that bounds all the $n$ dubits so that they will pass
through the upper slit as a whole or pass through the lower slit
as a whole. This is the function of QWD. The giant molecule is a
good candidate quantum system for such purpose. It has been
demonstrated that giant molecules exhibit quantum
interference\cite{r31} even though the internal states are
complicated. Here we use a hypothetical molecule which contains
many spin-1/2 nuclear spins, each  with different chemical shifts.
The molecule is placed in very low-temperature so that its
internal states are in the ground state. Assume that we are able
to manipulate single molecule and detect the single nuclear spin
state. So that we can also perform computing gate operations on
these nuclear spins and do so while they are flying slowly. We can
design a  duality computer as shown in Fig.\ref{f3}. A giant
molecule is moving slowly at extremely low temperature so that all
its internal states are in the ground state. We use the many
nuclear spins as the dubits. The molecule moves inside a tube. A
QWD is a two-branch junction where the molecule may move to either
the upper path or the lower path. Or we say that the wave function
of the molecule is divided into the upper and lower paths. A QWC
is simply the joining of the two tubes into a single tube. We
assume that the upper path and the lower path have the same
length, so the spatial wave function phase difference is zero.
When no gate operations
 are performed, the two sub-waves from the two
paths are always in phase to give constructive interference. The
device is placed in a appropriate strong static magnetic field. Gate
operations are performed using nuclear magnetic resonance(NMR)
technology.

\begin{figure}[h]
\begin{center}
\caption{The Walsch-Hadmard duality gate in GMDC.}
\includegraphics[width=10cm]{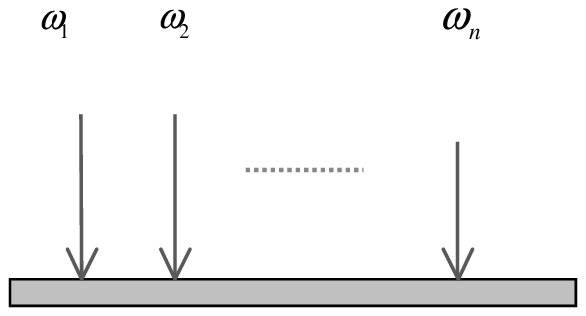}
 \label{f4}
\end{center}
\end{figure}

 Because the giant molecule is in constant motion, the
implementation of the gates is different from that of quantum computer using NMR. For
instance, if the nuclear spin is used as dubit, then different dubit has different
resonance frequency, and dubit can be addressed individually by the specific resonance
frequency.  As the speed of the molecule is very slow, relativistic effect can be
ignored. Suppose the molecule moves at a constant speed $v$. If a radio-frequency(rf)
pulse needed lasts for $t$ times long, then we can let the molecule move under such a
rf radiation for $L=vt$ long in the tube. Therefore in this scheme, a gate operation is
just like a molecule going through a series of "showers" of rf radiations and free
evolutions in the tube. These "showers" of gates can be placed at the single path
before the QWD just like that in a quantum computer where we need not to divide the
wave function. For instance Hadmard-Walsh gate operation
\begin{eqnarray}
|\psi_0\ket=|0\ket\equiv|0_10_2...0_n\ket\longrightarrow{1\over
\sqrt{2^n}}\sum_{i=0}^{2^n-1}|i\ket
\end{eqnarray}
can be performed by letting the molecule go through a section of
the tube with length $L=vt$ while releasing rf radiations with
resonance frequencies of all the dubits as shown in Fig.\ref{f4}.
 After the
molecule passes through the QWD, the quantum wave is divided into
two sub-waves travelling in different paths where different
unitary gate operations are performed. Thus if a molecule passes
through path $i$, its internal wave function will change to
\begin{eqnarray}
{1\over 2}|\psi\ket\rightarrow {1\over 2}U_i|\psi\ket,
\end{eqnarray}
where the subscript refers to the specific path. Then they
recombine to give the wave
\begin{eqnarray}
{(U_1+U_2)\over 2}|\psi\ket.
\end{eqnarray}
Then by measuring the internal state of the molecule, for instance using a device such
as a Stern-Gerlach apparatus, the computation result is read out.  In each path, the
computing gate operations can be any operation as that from quantum computation.

\subsection{Nonlinear quantum optics realization
scheme\cite{rabs}}

We use the polarization of a photon as the two states of a dubit. Photons with
different frequency are used to represent different dubits. Thus we need not to
consider the identical particle effect of quantum mechanics.

\subsubsection{Basic optical components}
First we describe some basic optical components: 1) polarization
beamsplitter; 2)the dichromic beamsplitter; 3) parametric down
conversion type I; 4) parametric down conversion type II; 5)
parametric up-conversion type I; 6) parametric up-conversion type
II. The scheme presented here is for the proof-of-principle
purpose. In
particular, the low efficiency of the up/down conversion is one of the
biggest difficulties in the implementation, however we assume them as 100\% as no theoretical
limit this possibility.

\begin{figure}[h]
\begin{center}
\caption{A polarization beamsplitter}
\includegraphics[width=10cm]{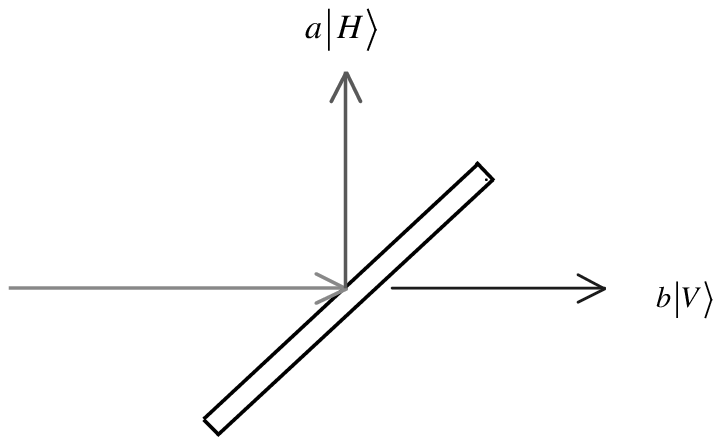}
 \label{f5}
\end{center}
\end{figure}

A polarization beamsplitter transmit the vertical polarized light
while reflect the horizontal polarized light. When a photon with
the following state
\begin{eqnarray}
|\psi\ket=a|H\ket+b|V\ket,
\end{eqnarray}
hits a polarization beamsplitter, the state changes to
\begin{eqnarray}
a|H\ket_1+b|V\ket_1\rightarrow a |H\ket_3+b|V\ket_2
\end{eqnarray}
as shown in Fig.\ref{f5}.

A dichroic beamsplitter is a device to separate or combine beams
of different wavelength. A long wave pass(LWP) dichroic
beamsplitter always transmit the longer wavelength wave and
reflects the shorter wavelength. A short wave pass(SWP) dichroic
beamsplitter always transmits the shorter wavelength and reflects
the longer wavelength. When a beam of light with both wavelength
$\lambda_1$ and $\lambda_2$ enters dichroic beamsplitter, one
wavelength is reflected and the other is transmitted. Then the two
beams are separated so that we can perform individual operations
on a specific wavelength photon, for instance a phase modulation,
or a polarization modulation. This is very convenient for duality computation.
When we
inverse the directions of the two beams in a dichroic beamsplitter,
we can combine the two
separated beams so that they become a single beam with different
wavelengths. We can use multiple dichroic beamsplitters to
separate out individual wavelength photon.  Dichroic beamsplitter
is different from QWD in the sense that a dichroic beamsplitter separates
two photons with different wavelengths into different paths whereas a QWD
divides the wave function for the whole two photons into two parts where in each path
the sub-wave describe the two photons.

\begin{figure}[h]
\begin{center}
\caption{Type I up-conversion. (a) Case 1 of type I SFG; (b) Case
2 of type I SFG.}
\includegraphics[width=10cm]{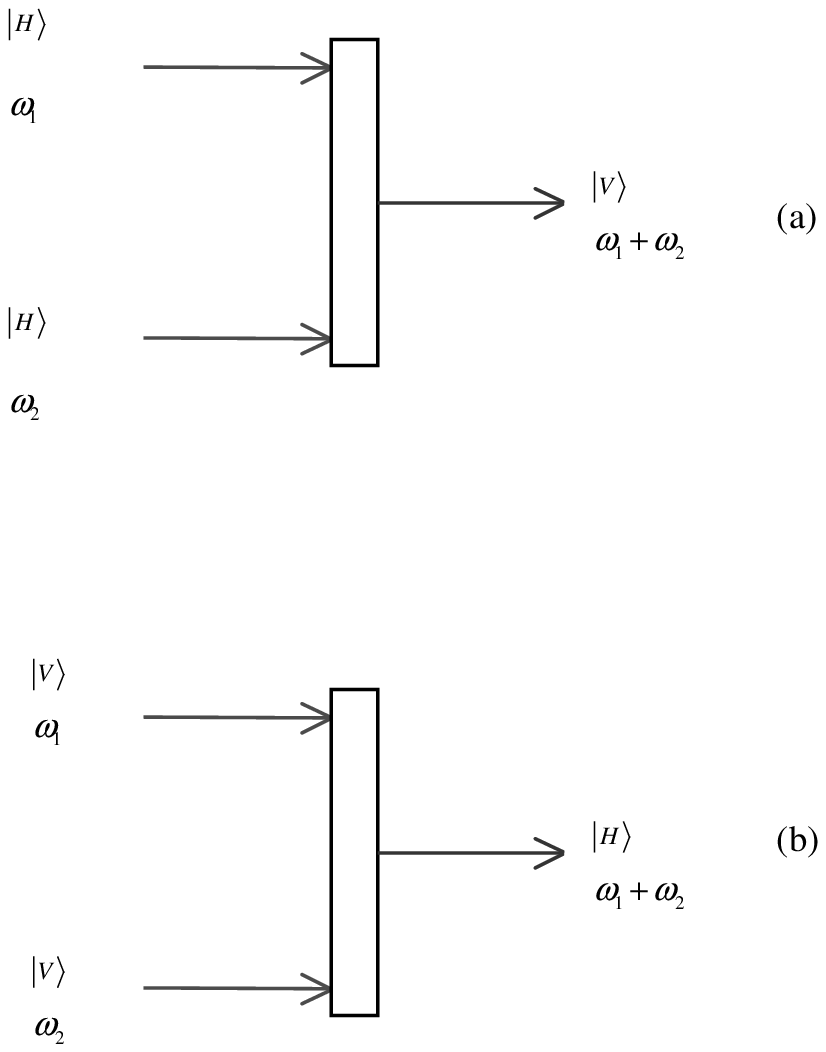}
 \label{f6}
\end{center}
\end{figure}

\begin{figure}[h]
\begin{center}
\caption{Type II up-conversion. (a) Case 1 of type II SFG; (b)
Case 2 of type II SFG.}
\includegraphics[width=10cm]{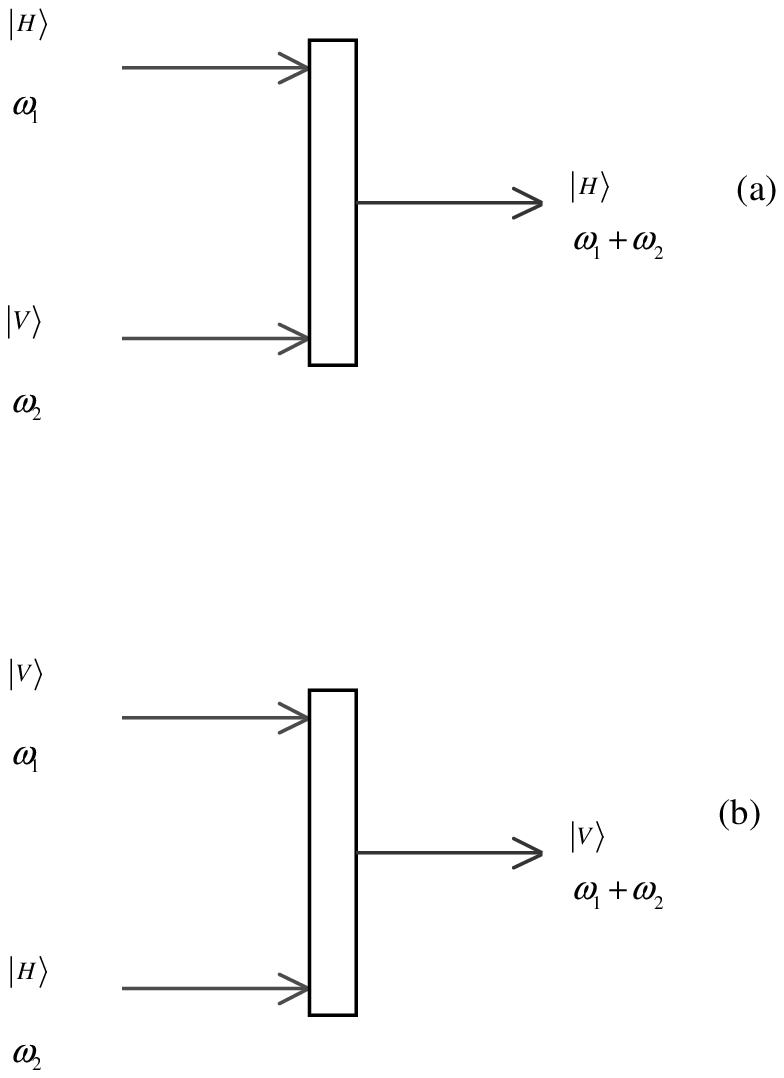}
 \label{f7}
\end{center}
\end{figure}

Sum frequency generation(SFG) is a process in which two photons
with frequency $\omega_1$ and $\omega_2$ respectively  are
combined to form a single photon with frequency
$\omega_1+\omega_2$ as shown in Fig.\ref{f6}. There are two types
according to the polarization states of the photons. In type-I
SFG, the process produces a photon with higher energy from two
photons with identical polarizations, whereas in type-II SFG, the
polarizations of the two input photons are different.
Mathematically, for type-I SFG, we have
\begin{eqnarray}
|H\ket_1|H\ket_2&&\rightarrow|V\ket_{12},\nonumber\\
|V\ket_1|V\ket_2&&\rightarrow|H\ket_{12},
\end{eqnarray}
as shown in Fig.\ref{f6}, whereas for two photons with arbitrary
polarizations, the change is
\begin{eqnarray}
(a_1|H\ket_1+b_1|V\ket_1)(a_2|H\ket_2+b_2|V\ket_2)\rightarrow
a_1a_2|H\ket_{12}+b_1b_2|V\ket_{12}.
\end{eqnarray}
For type-II SFG as shown in Fig.\ref{f7}, the changes are
\begin{eqnarray}
|H\ket_1|V\ket_2&&\rightarrow |H\ket_{12},\nonumber\\
|V\ket_1|H\ket_2&&\rightarrow |V\ket_{12},
\end{eqnarray}
and for two input photons with arbitrary polarization states, the
change is
\begin{eqnarray}
(a_1|H\ket_1+b_1|V\ket_1)(a_2|H\ket_2+b_2|V\ket_2)\rightarrow
a_1b_1|H\ket_{12}+b_1a_2|V\ket_{12}.
\end{eqnarray}
Here the subscript 12 means the up-converted photon.

\begin{figure}[h]
\begin{center}
\caption{Type I down-conversion: (a) Case 1; (b) Case 2.}
\includegraphics[width=10cm]{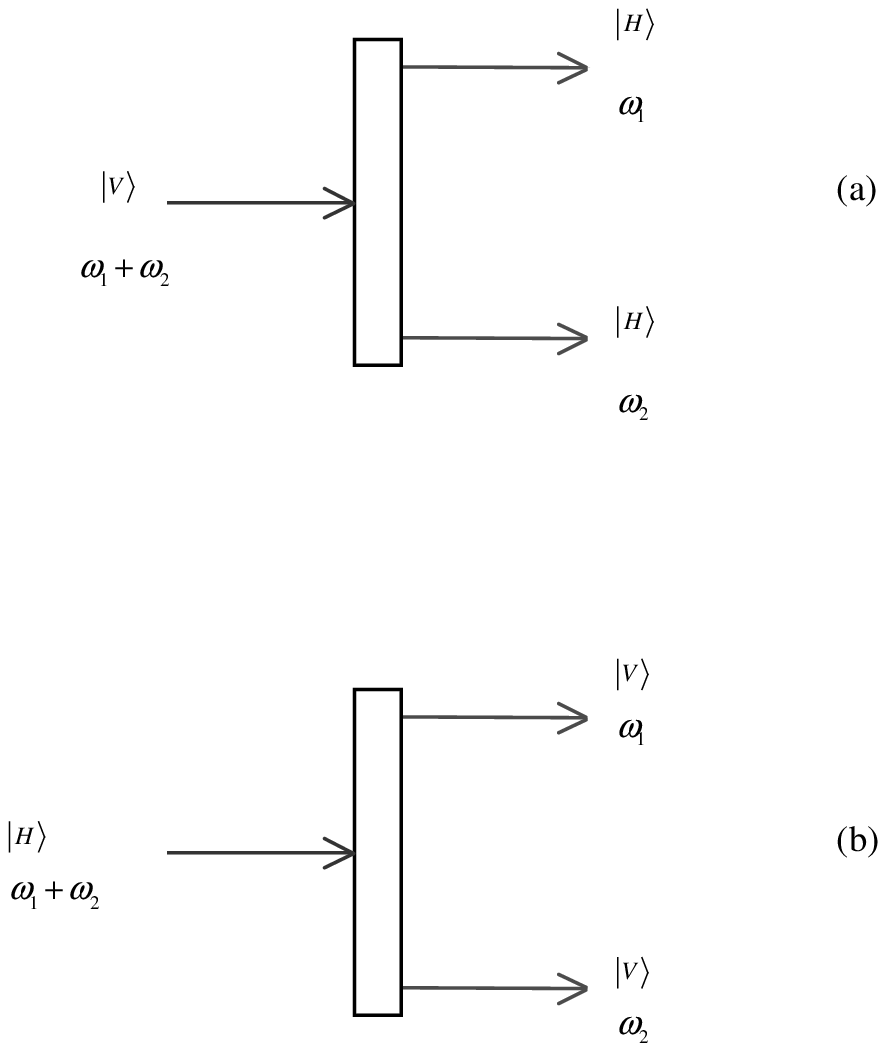}
 \label{f8}
\end{center}
\end{figure}

\begin{figure}[h]
\begin{center}
\caption{Type II down-conversion: (a) Case 1; (b) Case 2.}
\includegraphics[width=10cm]{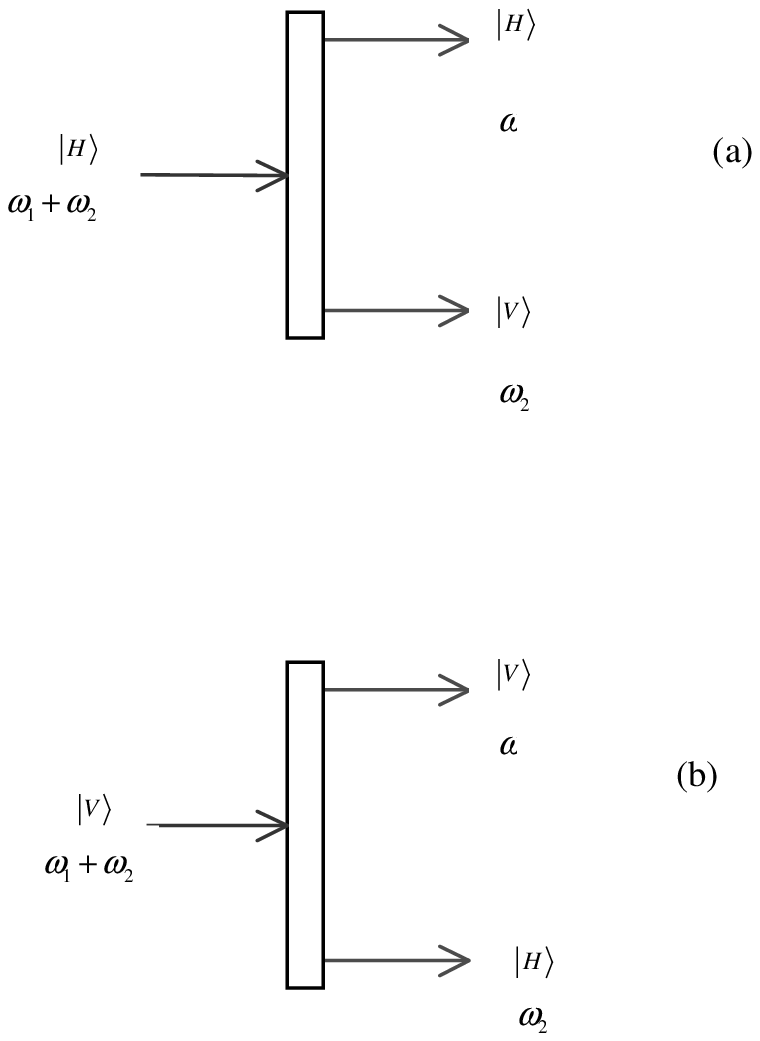}
 \label{f9}
\end{center}
\end{figure}

Spontaneous parametric down conversion(SPDC) is like the inverse
process of SFG and it generates two photons with frequencies
$\omega_1$ and $\omega_2$ on input of a single photon with
frequency $\omega_1+\omega_2$ as shown in Fig.\ref{f8}. There are
also two types of SPDC according to the polarizations of the
participating photons. In type-I SPDC, the two outgoing photons
have the same polarization states whereas in type-II SPDC, the two
outgoing photons have different polarization states. Specifically,
for type-I SPDC, the changes are
\begin{eqnarray}
|H\ket_{12}\rightarrow &&|V\ket_1|V\ket_2,\nonumber\\
|V\ket_{12}\rightarrow &&|H\ket_1|H\ket_2.
\end{eqnarray}
For type-II SPDC as shown in Fig.\ref{f9}, the changes are
\begin{eqnarray}
|H\ket_{12}\rightarrow && |H\ket_1|V\ket_2,\nonumber\\
|V\ket_{12}\rightarrow && |V\ket_1|H\ket_2.
\end{eqnarray}
When the input photon has an arbitrary polarization, the
corresponding changes will be the combined effects of both the
horizontal polarization and the vertical polarization components.
For instance, for type-I SPDC with input photon in state
$a|H\ket_{12}+b|V\ket_{12}$, the change is
\begin{eqnarray}
a|H\ket_{12}+b|V\ket_{12}\rightarrow
a|V\ket_1|V\ket_2+b|H\ket_1|H\ket_2.
\end{eqnarray}
Similar expressions can be written out explicitly for type-II
SPDC.

\subsubsection{Basic Unitary Computing Gates}

The single dubit gate operations are the phase rotation gate and
the rotation gate. In phase rotation gate, denoted by $P_\lambda$
the wave function acquires an arbitrary phase. Operating on a state
$a|H\ket+b|V\ket$, it changes the state according to
\begin{eqnarray}
P_\lambda|\psi\ket=e^{i\lambda}|\psi\ket=e^{i\lambda}(a|H\ket+b|V\ket).
\end{eqnarray}
This can simply be implemented with a variable wave plate.

 To rotate the dubit in the 2-dimensional space span by the two
 orthogonal polarizations, a polarization modulator is sufficient.
 It rotates the polarization of a photon through angle $\theta$, so
 a rotation operation $R_\theta$ makes the following changes
 \begin{eqnarray}
 R_\theta|H\ket&&=\cos\theta|H\ket+\sin\theta|V\ket,\nonumber\\
 R_\theta|V\ket&&=-\sin\theta|H\ket+\cos\theta|V\ket.
 \end{eqnarray}

\begin{figure}[h]
\begin{center}
\caption{The CNOT gate in  Nonlinear Quantum Optics DC.}
\includegraphics[width=10cm]{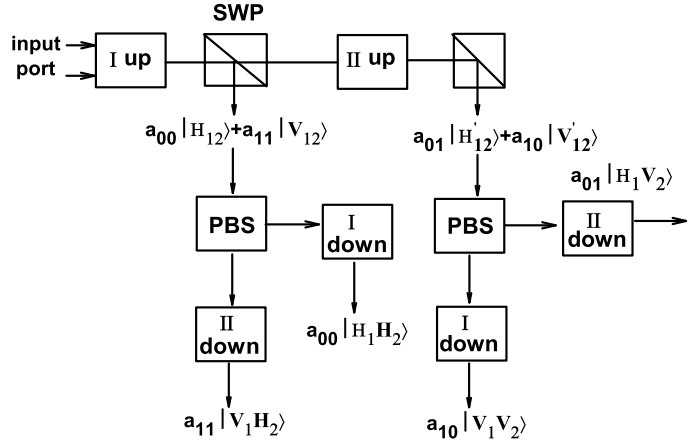}
 \label{f10}
\end{center}
\end{figure}

 The controlled NOT (CNOT)  gate is the most important gate
 operation in quantum computing and it is also an important gate
 operation in  duality computing. Here we propose a CNOT
 gate operation scheme using nonlinear quantum optics. It is a
 generalization of the method for full Bell-basis
 measurement used in Ref.\cite{r20}. It is shown in Fig.\ref{f10}. Two
 photons with frequencies $\omega_1$ and $\omega_2$ are two
 input dubits. They are in a most general superposition of their
 polarization basis states
 \begin{eqnarray}
 |\psi_{in}\ket=a_{00}|H_1H_2\ket+a_{01}|H_1V_2\ket+a_{10}|V_1H_2\ket+a_{11}|V_1V_1\ket.
 \end{eqnarray}
The effect of a CNOT gate is to perform an operation that changes
the state to
\begin{eqnarray}
 |\psi_{in}\ket=a_{00}|H_1H_2\ket+a_{01}|H_1V_2\ket+a_{10}|V_1V_2\ket+a_{11}|V_1H_1\ket,
 \end{eqnarray}
that is when the control dubit is in $1$(V), the state of the
target dubit is flipped. The nonlinear optics scheme works as
follows. First the two photons enters four summed frequency
generation(SFG) crystals to generate a single photon with the
summed frequency $\omega_1+\omega_2$. When they enter the type-I
SFG group, the crystal with optic axis aligned in the horizontal
direction combines the $|H_1H_2\ket$ component to form
$|V_{12}\ket$, and the crystal with vertical optic orientation
combines the $|V_1V_2\ket$ component to form $|H_{12}\ket$.
The
component $|H_1V_2\ket$ and $|V_1H_2\ket$ continue to propagate
into the type-II SFG crystal group and they combine to produce
$|H_{12}'\ket$ and $|V_{12}'\ket$ respectively. Dichroic
beamsplitters, polarization beamsplitters and mirrors are used to
separate the different components. Here the state has become
\begin{eqnarray}
a_{00}|V_{12}\ket+a_{01}|V'_{12}\ket+a_{10}|H'_{12}\ket+a_{11}|H_{12}\ket,
\end{eqnarray}
and thw different components are in different paths. To achieve the CNOT output state,
we need to use the SPDC process to generate two photons with frequencies $\omega_1$ and
$\omega_2$ respectively and leaving the $|H_1H_2\ket$ and $|H_{1}V_2\ket$ components
unchanged, but changes the component $|V_1H_2\ket$ and $|V_1V_2\ket$. This can be
completed by applying the type-I down conversions for $|V_{12}\ket$ which produces
component $|H_1H_2\ket$ and for $|H_{12}'\ket$ which produces $|V_1V_2\ket$ component
respectively. Similarly, the type-II SPDC processes for $|H_{12}\ket$ and
$|V'_{12}\ket$ produces $|H_1V_2\ket$ and $|V_1H_2\ket$ components respectively. Then
the different components are combined and put into two optical paths with frequency
$\omega_1$ and $\omega_2$ respectively. Hence the final state becomes
\begin{eqnarray}
a_{00}|H_1H_2\ket+a_{01}|H_1V_2\ket+a_{10}|V_1V_2\ket+|V_1H_2\ket,
\end{eqnarray}
the desired state.

\begin{figure}[h]
\begin{center}
\caption{Coherent generation of multi-dubits in the Nonlinear
Quantum Optics DC using cascade down-conversions}
\includegraphics[width=10cm]{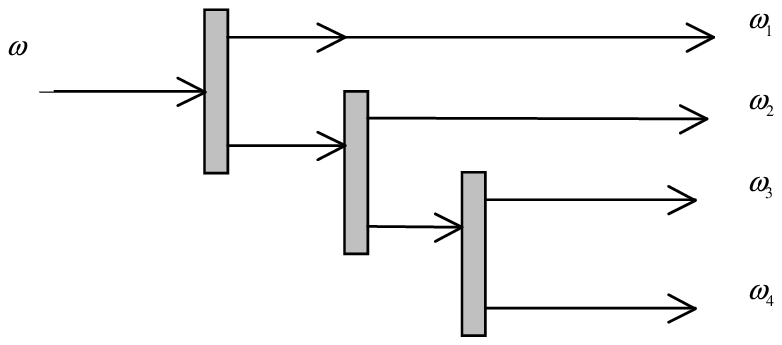}
 \label{f11}
\end{center}
\end{figure}

The most difficult part is the QWD. We cannot simply let $n$ dubits to pass through a
beamsplitter to produce a division of the wave, since there will be $2^n$ possibilities
as has been pointed out in the preceding section. We need that the $n$ dubits system as
a whole be divided into two sub-waves. To achieve this, we need the $n$ dubit photons
be generated simultaneously and coherently. First we look at how to produce $n$ dubit
photons coherently using SPDC. It can be realized in the setup as shown in
Fig.\ref{f11}. Suppose we have a photon with frequency
$\omega=\omega_1+\omega_2+...+\omega_n$ and is in state $|V\ket$. Then we let this
photon to pass through a type-I SPDC crystal so that it produces two photons in state
$|H_1H_1'\ket$ where the first photon has frequency $\omega_1$ and the second photon
has frequency $\omega_1'=\omega_2+...+\omega_n$. Then the second photon continues to
pass through a type-II SPDC crystal to produce two photons in state $|H_2V_2'\ket$,
with the second photon having frequency $\omega_2'=\omega_3+...+\omega_n$. After $n-1$
such SPDC processes, we produce $n$ photons of different frequencies. These $n$ photons
can be adjusted so that they arrive in a plane perpendicular to their propagation at
the same time. Unitary gate computations can be performed on them using the a series of
single dubit operation and the CNOT gate.

\begin{figure}[h]
\begin{center}
\caption{A QWD in Nonlinear Quantum Optics DC. The 4-dubits in the
upper path and lower path have identical polarization states. }
\includegraphics[width=10cm]{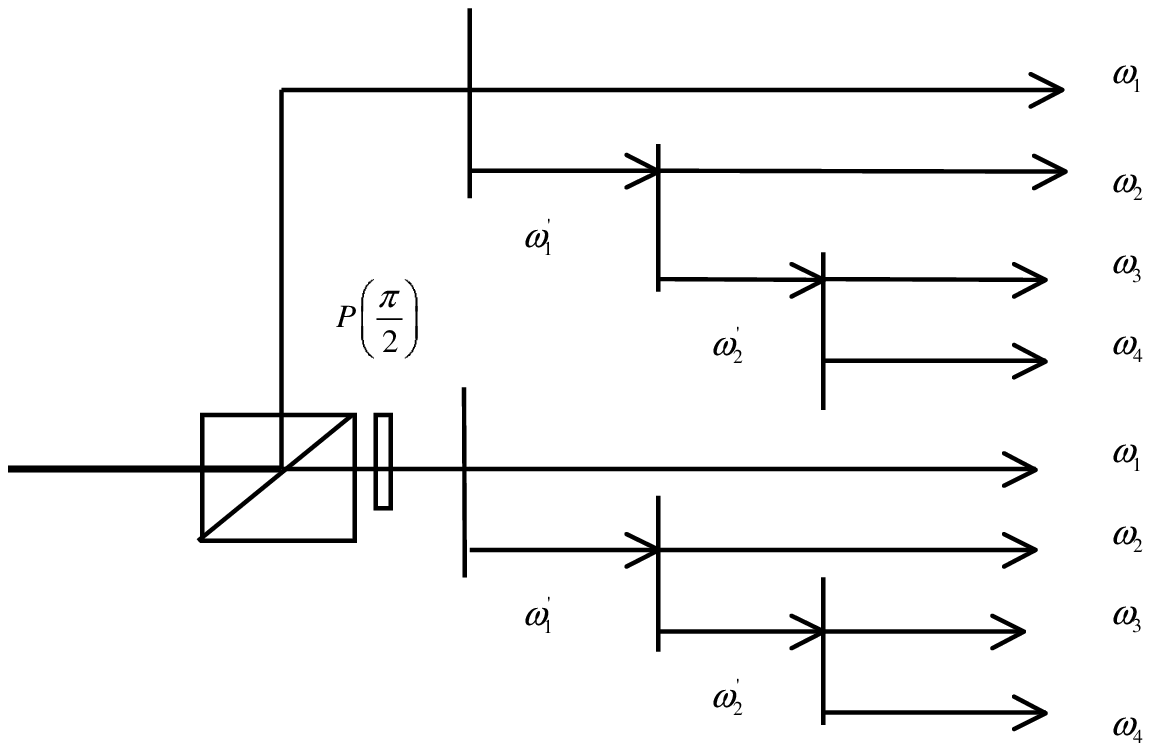}
 \label{f12}
\end{center}
\end{figure}

Instead of making this $n$ photon wave to divide into two
sub-waves with identical internal polarization state, we perform a
quantum wave
 division at the
source: we use a beamsplitter before the first SPDC crystal as
shown in Fig.\ref{f12}, then the wave function of the photon with
 $\omega=\omega_1+\omega_2+...+\omega_n$
is divided into sub-waves. A half-wavelength plate is used to
compensate the phase difference between transmitted and reflected waves.
Each of the sub-wave will further be transformed into a sub-wave of
$n$ photons with the coherent photon production process described above.
 Then the sub-wave in the upper path  is
\begin{eqnarray}
{1\over \sqrt{2}}|H_1H_2...H_n\ket_u,
\end{eqnarray}
whereas in the lower path the sub-wave is
\begin{eqnarray}
{1\over \sqrt{2}}|H_1H_2...H_n\ket_d.
\end{eqnarray}

\begin{figure}[h]
\begin{center}
\caption{A Global View of Nonlinear Quantum Optics DC.}
\includegraphics[width=10cm]{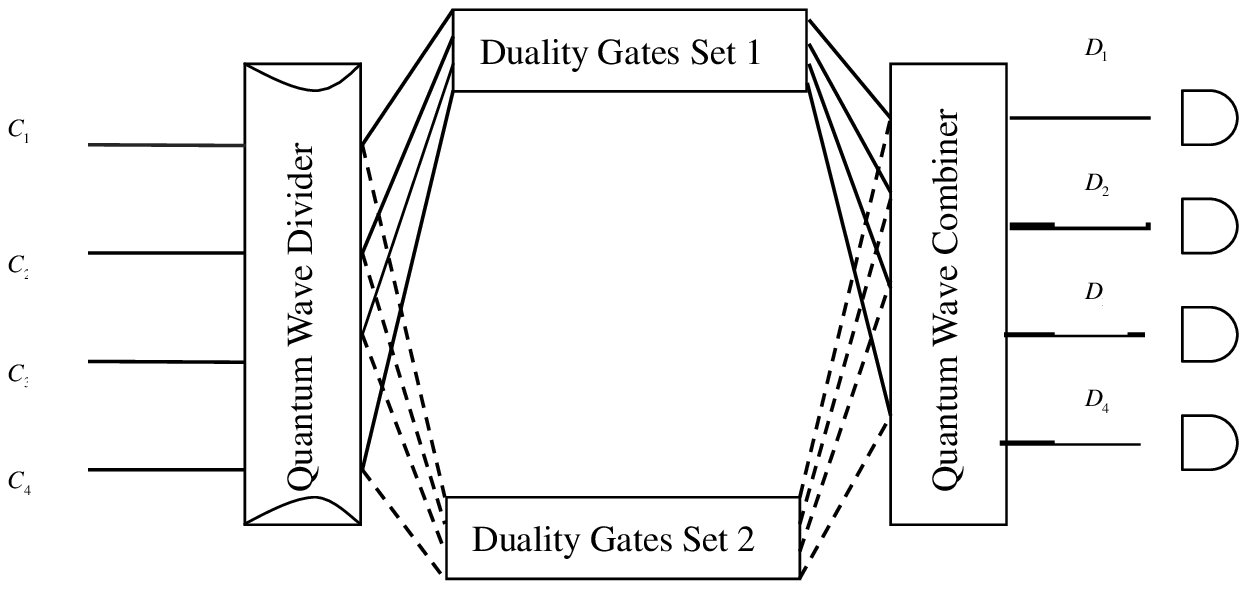}
 \label{f13}
\end{center}
\end{figure}

The whole  duality computing process can be shown in
Fig.\ref{f13}. The computing unitary gate operations are performed
on the upper and lower paths respectively. Then the two path waves
are joined finally at the QWC. Because the two paths have the same
optical path, the two waves should add up together, therefore the
wave after the QWC is
\begin{eqnarray}
{1\over 2}\sum_{i=0}^{2^n-1}(c_{i,u}+c_{i,d})|i\ket.
\end{eqnarray}

A QWC can be easily constructed by directing the specific photon
from the upper path and the lower path so that they coincide in
space and time, for instance as in Ref.\cite{independent}.

\begin{figure}[h]
\begin{center}
\caption{The read-out device in the NODC.}
\includegraphics[width=10cm]{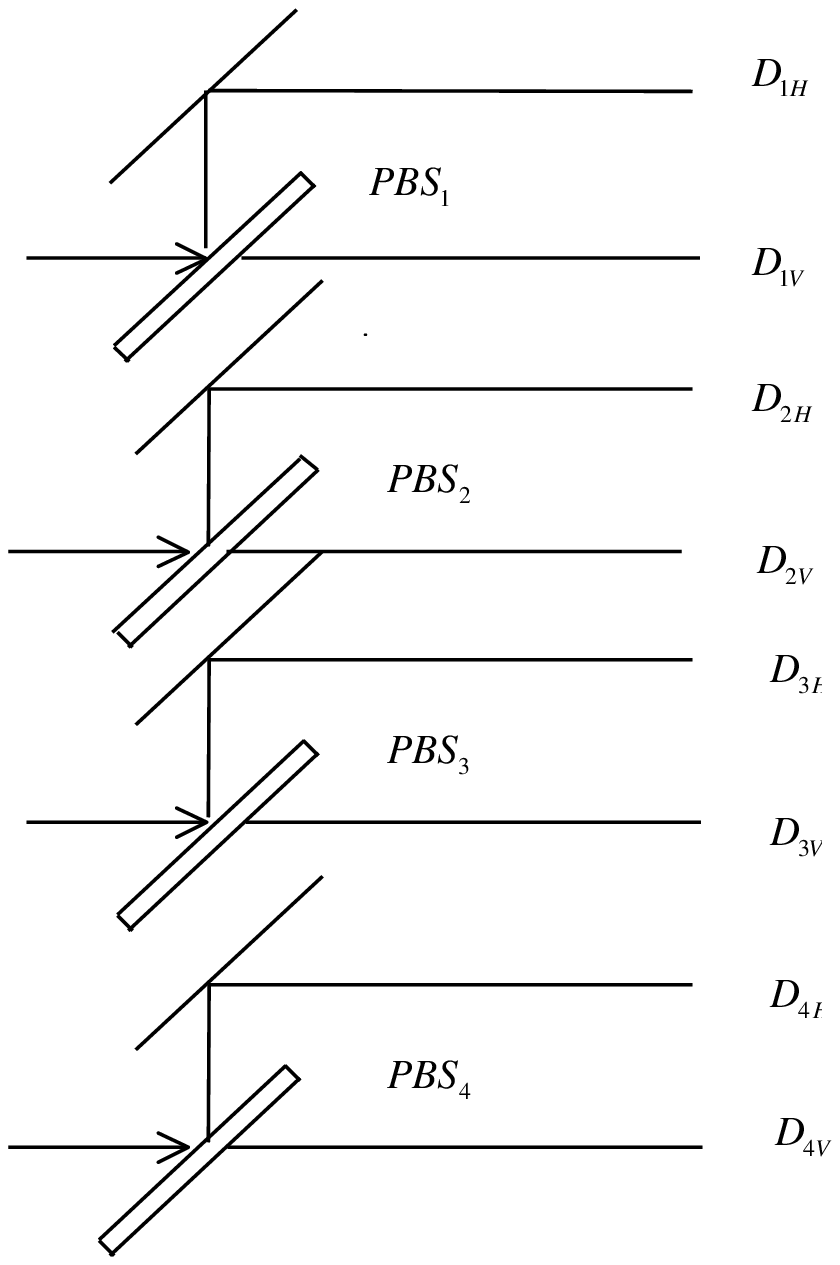}
 \label{f14}
\end{center}
\end{figure}

The measurement device is constructed as shown in Fig.\ref{f14}.
After the QWD, each dubit is guided to a polarization beamsplitter
where the horizontal and vertical components are split and they
further travel to two  detectors.  One of the two detectors will
register a click which is a projective measurement for the dubit.
We can read out the state of a photon by looking at which detector
clicks. If the upper detector clicks, it corresponds to 0, and if
the lower detector clicks it corresponds to state 1. Altogether
there are $2n$ detectors, and they corresponds to the bit value of
the $n$ dubits. This measurement is a projective measurement. In
this proof-of-principle design, I assume that all the components
are ideal and have 100\% efficiency. The up and down conversions
are also of 100\% efficient.

\section{A Single Query Unsorted Database Search and Polynomial Algorithm
for NP-complete Problems in Duality Computer} \label{s7}

We first present an algorithm that may find an marked item from an unsorted database
with a single query. This is an exponentially fast search algorithm. Then we show that
NP-complete problem can be reduced to unsorted database search, hence NP-complete
problem may become P problem in duality computer. However, the validity of the
algorithm depends on the efficiency of measurement for a partial wave function which
will be discussed shortly.

\subsection{A duality computer algorithm for unsorted database
search}

The unsorted database search problem is a benchmark  for computing. To find a marked
item from an unsorted database of $N$ items, a classical computer requires $O(N)$
steps. A quantum computer requires $O(\sqrt{N})$ steps using the Grover
algorithm\cite{r25} or an improved algorithm with 100\% successful rate\cite{r26}. By
introducing classical parallelism, the problem can be further speeded up modestly.
Using the Br\"{u}schweiler algorithm\cite{brusch}, $O(\ln N)$ steps is required to find
the marked state. Using the Xiao-Long algorithm, one needs only a single query to find
the marked state\cite{xiaolong}. However this speedup is achieved at the cost of more
computing resources. Namely now there are $O(N)$ quantum computers working in parallel.
In general with $N_2$ quantum computers working in parallel, the number of queries
required to find the marked state is $O(\sqrt{N/N_2})$ in a parallelized quantum
computing\cite{parallel}. Though these algorithms achieve speedup by using more
resources, they are still useful in ensemble quantum computers such as liquid nuclear
magnetic resonance \cite{7qubit}.

In duality computer, the speedup may be achieved without the use
of additional resource. An algorithm is presented as follows.

1. Prepare the state of the duality computer in the equally
distributed state,
\begin{eqnarray}
|\psi_0\ket=\sqrt{1\over N}\sum_{i=0}^{N-1}|i\ket=\sqrt{1\over N}
(|0\ket+|1\ket+...+|\tau\ket+...+|N-1\ket),\label{even}
\end{eqnarray}
where $\tau$ is the marked item we are searching.

2.  Let the duality computer go through QWD, so that it divides
the wave into two sub-waves
\begin{eqnarray}
|\psi_u\ket&=&{1\over 2\sqrt{N}}\sum_{i=0}^{N-1}|i\ket={1\over
2\sqrt{N}}(|0\ket+|1\ket+...+|\tau\ket+...+|N-1\ket),\label{upper}\\
|\psi_d\ket&=&{1\over 2\sqrt{N}}\sum_{i=0}^{N-1}|i\ket={1\over
2\sqrt{N}}
(|0\ket+|1\ket+...+|\tau\ket+...+|N-1\ket)\label{lower}.
\end{eqnarray}

3. Apply the query to the lower-path sub-wave, reverse the
coefficients of all basis states $|i\ket$ except the marked state
$|\tau\ket$, the lower sub-wave becomes
\begin{eqnarray}
|\psi_d\ket'&=&{1\over 2\sqrt{N}}
(-|0\ket-|1\ket+...+|\tau\ket-...-|N-1\ket).
\end{eqnarray}

No operation is applied to the upper-path sub-wave, it remains in
state as in Eq. (\ref{upper}).

4. Combine the sub-waves at the QWC, and the wave becomes
\begin{eqnarray}
|\psi_f\ket=\sqrt{1\over N} |\tau\ket.  \label{final}
\end{eqnarray}

5. Make a read-out measurement, and the marked item $\tau$ will be found with some probability
depending on further studies on the measurement of a partial wave function as will be
discussed later. In a best
case scenario, the measurement gives the marked item with certainty.

 The query can be implemented using $O(\ln N)$ number of dubits. As
the size of the database $N$ increases, the difficulty in
constructing the query increases only logarithmically.

\subsection{Duality algorithm for NP-Complete Problems}

According to the Cook-Levin theorem, all NP-complete problems are
polynomially equivalent. One can reduce one NP-complete problem in
polynomially equivalent steps into another NP-complete problem.
Let's look at the SATISFIABILITY problem, where one needs to find
if there are solutions to a logical expression,
\begin{eqnarray}
P(x_1,x_2,...x_n)=1,\label{sat}
\end{eqnarray}
of $n$ binary variables.  To find its solution, we cast the
problem into an unsorted database search problem, though the
reverse is not possible in general. The algorithm is as follows

1. Prepare the state of the duality computer in the equally
distributed state,
\begin{eqnarray}
|\psi_0\ket=\sqrt{1\over N}\sum_{i=0}^{N-1}|i\ket=\sqrt{1\over N}
(|0\ket+|1\ket+...+|\tau_1\ket++|\tau_2\ket...+|N-1\ket),
\end{eqnarray}
where $\tau_i$ is one of the item that satisfies the logical
equation Eq.(\ref{sat}).

2.  Let the duality computer go through QWD, so that it divides
the wave into two sub-waves
\begin{eqnarray}
|\psi_u\ket&=&{1\over 2\sqrt{N}}\sum_{i=0}^{N-1}|i\ket={1\over
2\sqrt{N}}(|0\ket+|1\ket+...+|\tau_1\ket++|\tau_2\ket+...+|N-1\ket,\label{upper2}\\
|\psi_d\ket&=&{1\over 2\sqrt{N}}\sum_{i=0}^{N-1}|i\ket={1\over
2\sqrt{N}}
(|0\ket+|1\ket+...+|\tau_1\ket++|\tau_2\ket+...+|N-1\ket)\label{lower2}.
\end{eqnarray}

3. Using the logical expression Eq. (\ref{sat}) as a query. This
can be implemented by using $O(n)$ number of dubits in polynomial
steps. Apply the query to the lower-path sub-wave, reverse the
coefficients of all basis states $|i\ket$ except those that
satisfy Eq.(\ref{sat}), the lower sub-wave becomes
\begin{eqnarray}
|\psi_d\ket&=&{1\over 2\sqrt{N}}
(-|0\ket-|1\ket-...+|\tau_1\ket+|\tau_2\ket+...-|N-1\ket).
\end{eqnarray}

No operation is applied to the upper-path sub-wave, it remains in
state Eq. (\ref{upper2}).

4. Combine the sub-waves at the QWC, and the wave becomes
\begin{eqnarray}
|\psi_f\ket=\sum_{i=1}^m\sqrt{1\over m}|\tau_i\ket,\label{final2}
\end{eqnarray}
where $m$ is the number of items that satisfies the logical
expression Eq.(\ref{sat}).

5. Make a read-out measurement, if Eq.(\ref{sat}) is satisfiable
one of the marked item $\tau_{i1}$ is found out. If it is not
satisfiable, no state is obtained.

6. To find all the solutions to the logical expression, one repeat steps 1, 2, 3, 4.
Now we let the wave in Eq. (\ref{final2}) go through a QWD again and divide it into two
paths. We inverse the state $|\tau_{i1}\ket$ in one path and then we combine them in a
QWC. Then the $|\tau_{i1}\ket$ component is deleted from the state Eq.(\ref{final2}).
Another measurement reads out a solution other than $\tau_{i1}$. Continuing in this
way, all the $m$ solutions to the problem are found.

\subsection{The measurement efficiency for a partial wave function}

It is worth mentioning the coefficient in Eq. (\ref{final}).  It can not simply be
renormalized to 1. For if the two sub-waves have opposite signs, they will cancel each
so that the probability will be zero which corresponds to a dark point in an
interference pattern. If the two sub-waves are in phase, then the probability of
detection will be the greatest and it will correspond to a bright spot in an
interference pattern. Then one would be expected to give the probability of $1/N$ for
obtaining the marked state from state in Eq. (\ref{final}) and this would make the
search algorithm useless. However, it is not so simple. It depends on the
interpretation of the wave function and the measurement in quantum mechanics. We
suggest three possibilities, and the final answer depends on further experiment.

First we point that the probability of finding state $|\tau\ket$ in
a state in Eq. (\ref{final})
and in state (\ref{even}) may be different. In Eq. (\ref{final}), the
situation is whether we
get a result for a measurement or do not get a result at all, while in Eq. (\ref{even}), we
always get a result which is one of the $N$ basis states, and $\tau$ will have only $1/N$
probability to be the result.

We give three possibilities:

1) The possibility is $1/N$ and there is no difference between Eq. (\ref{final}) and
Eq. (\ref{even}). This is true for some interpretations of quantum mechanics,
for instance in the ignorance interpretation where the particle is actually in some point
in space and the probability is because we do not know this knowledge.

2) The probability is 100\%, however one has to wait longer time to get a result. This
may be reasonable as the measuring process may need some interaction between the
duality computer system and the detector. The smaller the wave function exposed to the
detector, the longer it takes to get a result.

3) The probability is 100\% and it takes about the same period of time to get the
result. But it may require that the magnitude of the amplitude be greater than a
minimum value. Under the minimum value, the detector will not register a click, hence
no particle will be detected when the two-sub-waves are in total cancellation. However
when it is bigger than the minimum amount, the detector will register a click as if the
whole wave function is exposed to the detector. This can be understood in the giant
molecule scheme. If the two sub-waves have opposite signs so that total cancellation
takes place, it means that the giant molecule is trapped inside the tube as a standing
wave. If partial cancellation occurs, for instance in Eq.(\ref{final}), the major part
of the wave function is inside the tube as standing wave, and only a small part is
exposed to the detector. However here we just make measurement on this part of the wave
function, the other part of the wave function is not measured and they can not
collapse. Hence if no absorption takes place inside the tube, the detector always
obtains a result. If detectors were placed everywhere inside the tube, then the
detector would have $1/N$ probability to get a result as it is competing with other
detectors. However as no competing detectors are present, the detector at the final
point will always obtain a result in normal time span. For instance if we have a
balloon, and we use a board full of needles to punch it. Any one of the needles can
punch the balloon so that collapse it. However we can break the balloon with the same
ease if we just use one needle to punch the balloon in just a small area.

As a test, we propose the following thought experiment that may
also be performed practically. One traps a single electron in a
Penning trap. Then putting different number of detectors inside
the Penning trap such as that in Ref.\cite{penning}, and switch on
the detectors study the time period when one of the detectors
registers a click. The more the detectors, the larger the electron
wave function being measured.

We leave the problem open at this stage, and a detailed discussion will be published
elsewhere. For use in computing, the third possibility is preferred, and if it is true
duality computer will be the most powerful computing machine so far. It will be
interesting to study if it is possible to improve the efficiency by repetition or by
placing more detectors at other spatial phase difference points, or using the quantum
Zeno effect, in the worst scenario.

However, it should be emphasized that irrespective of the result on the measurement
efficiency, the duality computer can be made to work as a quantum computer: the upper
and lower sub-waves experience the same gate operation. Thus the computing power of
duality computer is at least as powerful as the quantum computer, even in the worst
case of the measurement efficiency.

\section{Concluding Remarks}
\label{s8}

We have proposed a general principle of quantum interference.
Whenever the sub-waves of a quantum system, whether single, bound
or loose, coincide in space and time, interference may occur.

The quantum computer uses only the particle nature of quantum
system, and it does not use the full power of quantum mechanics.

We propose the  duality computer. Quantum wave can be divided and
recombine. Different computing gate operations can be performed at
the different paths. This enables us to perform computation using
not only products of unitary operations, but also linear
combinations of unitary operations. This provides duality
parallelism which may outperform quantum parallelism in quantum
computer.

Two conceptual designs of duality computer are proposed. The giant
molecule scheme and the nonlinear quantum optics scheme.

Searching a marked item from an unsorted database may require only a
single query in duality computer. This is the holy grail of
unsorted database search problem.

NP-complete problem can be cast into an unsorted database search
problem. Thus NP-complete problem may become P problem in duality
computer. This may be the first possible computing machine to answer the
question if NP-complete is equivalent to P.

Open problem exists on the detection efficiency in the duality computer.
Three possibilities are proposed which are closely related to fundamental
issues in quantum mechanics. The final answer depends on further
experimental result, though a specific answer is favored for duality computing.

Building a duality computer maybe difficult. However taking into
account the enormous power it may provide, experimental endeavors are
worthwhile, in particular the test of the general interference principle for
complicated quantum systems and to find out the measurement efficiency of
a partial wave function.

This work is supported by the National Fundamental Research
Program Grant No. 001CB309308, China National Natural Science
Foundation Grant No. 10325521, 60433050, and the SRFDP program of
Education Ministry of China.

It should be noted that quantum interference also exists in quantum computer, for
instance in the Shor algorithm where wanted components are strengthened and unwanted
components are suppressed. However, in the quantum computer, unitarity of gate
operations are always obeyed. The quantum computer system as a whole exhibits particle
nature. The role of quantum interference in quantum computing has been studied by Ekert
\cite{ekert}, Galindo and Martin-Delgado \cite{martin}, Shiekh\cite{shiekh} and
Finkelstein\cite{shiekh}.

Caufield and Shamir proposed an optical computer employing the particle and wave
duality nature of light where one or several coherent light sources illuminate an
optical system containing the input variables and a detector array that records the
outputs\cite{wpd1}. Our model is different from their model in that the particle wave
duality nature is exhibited in a different manner. In the duality computer, the whole
quantum system of the computer system exhibits the particle wave duality nature whereas
inthe Caufield and Shamir computer model the computer system as a whole does not
exhibit the particle wave duality: it is always present there as a particle, and the
particle wave duality nature is only exhibited by the individual lights, thus it may be
a hybrid computer model of classical computer and quantum computer.

\end{document}